\documentclass[12pt]{article}
\usepackage{amsmath}
\usepackage{graphicx}
\usepackage{enumerate}
\usepackage{natbib}
\usepackage{url} % not crucial - just used below for the URL 

\newcommand{\blind}{1}

\usepackage{changes}
\usepackage{lipsum}
\definechangesauthor[name ={petros}, color = red]{PD}
\definechangesauthor[name ={mattia}, color = orange]{MS}
\definechangesauthor[name ={modification}, color = green]{mod}

\begin{document}

\def\spacingset#1{\renewcommand{\baselinestretch}%
{#1}\small\normalsize} \spacingset{1}

%%%%%%%%%%%%%%%%%%%%%%%%%%%%%%%%%%%%%%%%%%%%%%%%%%%%%%%%%%%%%%%%%%%%%%%%%%%%%%

\if1\blind
{
  \title{\bf Doubly-online changepoint detection for monitoring health status during sports activities}
  \author{Mattia Stival \thanks{
    The authors gratefully acknowledge the cardiologist Dr Costas Thomopoulos for valuable discussions. This research was supported by funding from the University of Padova Research Grant BIRD203991.
 %\textit{please remember to list all relevant funding sources in the unblinded version}
 }\hspace{.2cm}\\
    Department of Statistical Sciences, University of Padova\\
    and \\
    Mauro Bernardi \\
    Department of Statistical Sciences, University of Padova\\
    and \\
    Petros Dellaportas \\
    Department of Statistical Science, University College London  \\
    Department of Statistics, Athens University of Economics and Business  \\
    The Alan Turing Institute \\
  }
  \maketitle
} \fi

\if0\blind
{
  \bigskip
  \bigskip
  \bigskip
  \begin{center}
    {\LARGE\bf Doubly-online changepoint detection for monitoring health status during sports activities}
\end{center}
  \medskip
} \fi

\bigskip
\begin{abstract}
We provide an online framework for analyzing data recorded by smart watches during running activities. In particular, we focus on identifying variations in the behavior of one or more measurements caused by changes in physical condition, such as physical discomfort, periods of prolonged de-training, or even the malfunction of measuring devices. Our framework considers data as a sequence of running activities represented by multivariate time series of  physical and biometric data.
We combine classical changepoint detection models with an unknown number of components with Gaussian state space models to detect distributional changes between a sequence of activities. The model considers multiple sources of dependence due to the sequential nature of subsequent activities, the autocorrelation structure within each activity, and the contemporaneous dependence between different variables. We provide an online Expectation-Maximization (EM) algorithm involving a sequential Monte Carlo (SMC) approximation of changepoint predicted probabilities. As a byproduct of our model assumptions, our proposed approach processes sequences of multivariate time series in a doubly-online framework. While classical changepoint models detect changes between subsequent activities, the state space framework coupled with the online EM algorithm provides the additional benefit of estimating the real-time probability that a current activity is a changepoint.
\end{abstract}

\noindent%
\vfill

\newpage
\spacingset{1.9} % DON'T change the spacing!
\section{Introduction}
\label{sec:intro}
Running is one of the most popular and practiced sports worldwide, with almost $60$ million people having participated in running, jogging, and trail running in $2017$ in the United States \citep{statistarunning}. 
Increasingly more runners use smart watches and devices that record their workouts, allowing for performance analysis and the planning of future workouts.  Worldwide smart watch shipments volume as estimated by \citet{statistawearables} were $74$ million units in $2018$,  $97$ million units in $2019$, $115$ million units in $2020$, with an expected growth to over $258$ million units by $2025$. Apps and wearables are driving the next digital health and fitness revolution, in which intelligent and automatic real-time control and monitoring tools will become extremely relevant \citep{statistafitness}.  Indeed, it is expected that in the near future, smart watches may be used as medical monitoring devices, providing support at an individual level to health-care consumers \citep{free2013effectiveness, singh2018heart} and, more importantly, to users with different levels of health literacy, communication, and data skills \citep{do2019behavior, Vitabile2019}.  The spectrum of available and potential measurements by smart watches includes information on movement, heart rate, blood oxygenation and pressure, and glucose \citep{garcia2020microneedle, pkvitality}.   Our contribution provides a modeling framework to analyze, in an online fashion, data recorded from smart devices during running activities.  In particular, we focus on identifying variations in the behavior of one or more measurements caused by changes in physical condition such as physical discomfort, periods of prolonged de-training, or even the malfunction of measuring devices \citep[][]{schneider2018heart}.

The use of wearable technologies and sensor data for medical problems is gaining increasing interest from the statistical community, see for example  \citet[][]{multilevel_matvar, mixture_accelerometer, biometrika_mobile}.  
The difficulty in monitoring performances due to the presence of disturbing factors, such as environmental conditions or other within-activity sources of variability, is widely accepted; see, for example  \citet{schneider2018heart}. 
A valuable contribution to this field was provided by \cite{frick.kosmidis.17}, who developed an R \citep{Rstat} package 
that allows for both basic and advanced retrospective analysis of data collected from smart devices.  Unlike previous works on this type of data, we focus on online inference because it highlights the important aspect of smart devices related to the monitoring activities as they are carried out \citep{bourdon2017monitoring}. 

\begin{figure}
\begin{center}
    \includegraphics[scale = .25]{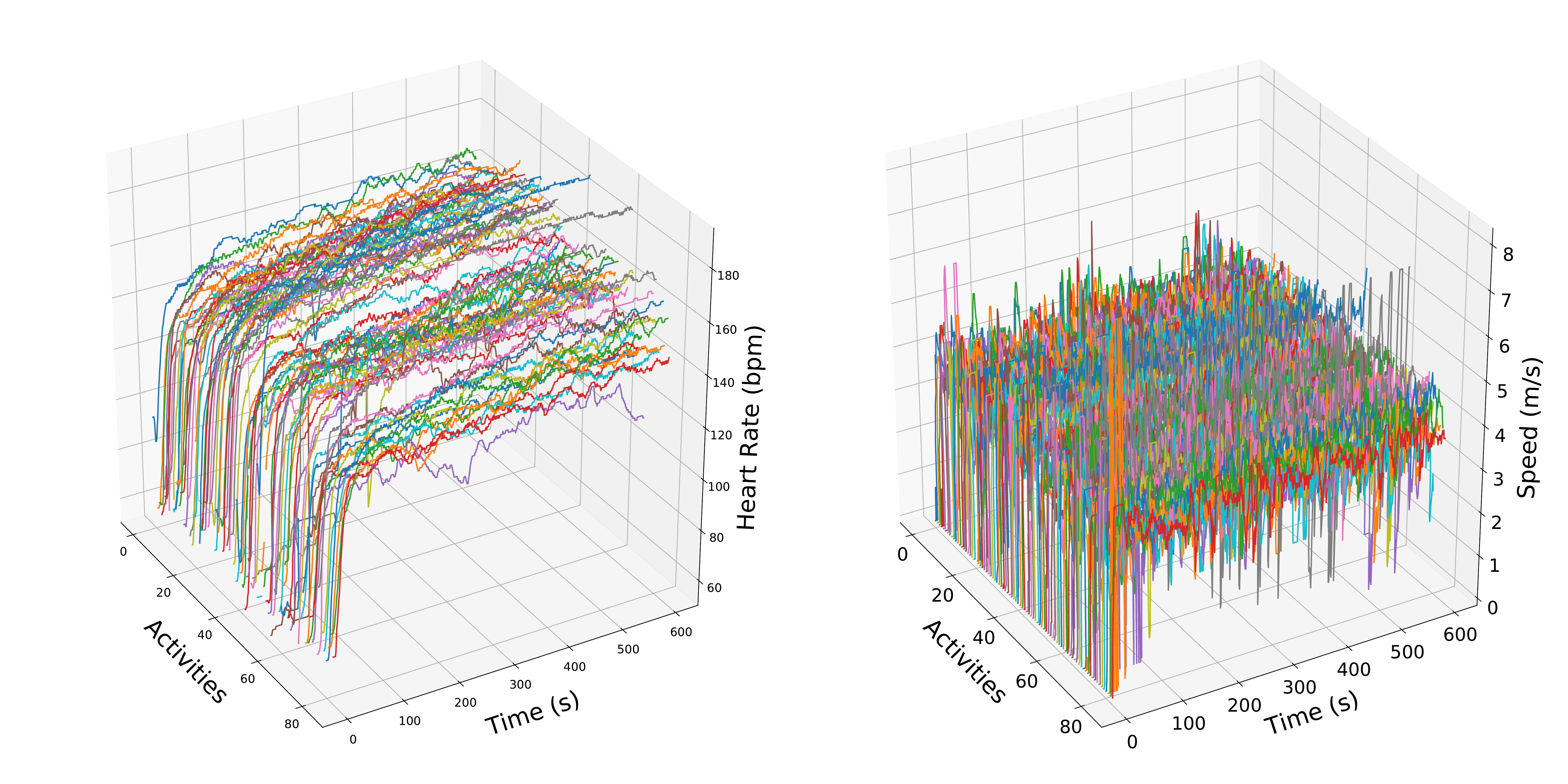}
    \caption{A sequence of activities performed by one athlete from our dataset}
    \label{fig:8act_no}
\end{center}
\end{figure}

Recent literature in sports science and medicine points out the need to make decisions by evaluating the personal medical history, the long- and short-term training goals of the athlete, and the time course of training schedules \citep{pelliccia20212020, schneider2018heart}. We address these issues by utilizing data collected as a sequence of {\em activities}, where each activity represents a part of the training session.  The relevant measurements that we will consider in this study are heart rate (bpm, beats per minute) and speed (m/s, meters per second), whereas other common variables that can be incorporated in our proposed methodology are cadence (spm, steps per minute) and the runner's geographical position (latitude, longitude, and altitude). Figure \ref{fig:8act_no} shows a sample of the data, consisting of $85$ consecutive warm-up activities performed by one athlete during which the heart rate and speed are monitored over time. For all the activities, after a sudden increase, the heart rate curves seem to slowly evolve around a trend, while the speed levels change slowly during the activity. 

\begin{figure}
\centering
\includegraphics[scale = 0.28]{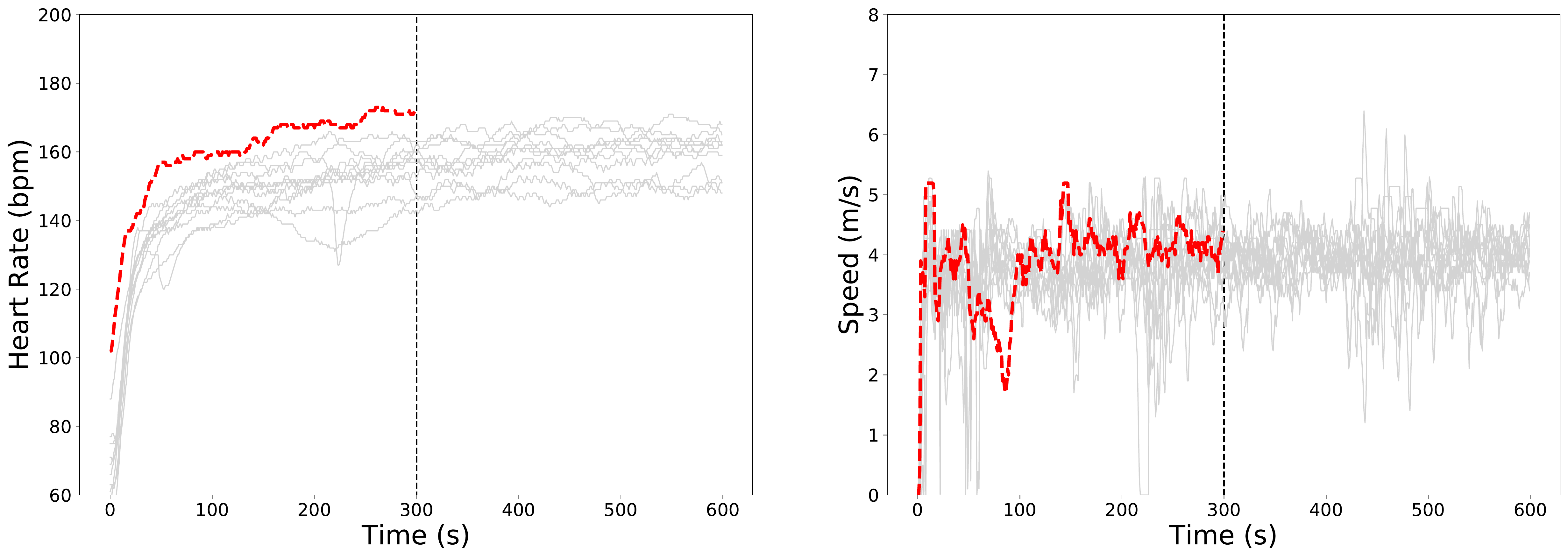}
\caption{An example of the within-online setting. The red dashed lines indicate the current monitored activity, while gray lines denote previous activities. The vertical line marks the time at which our algorithm provides the posterior probability that the current activity is a changepoint. }\label{fig:sketch}
\end{figure}

For one activity, all collected information is represented by a multivariate time series, with complex dependence structures that make the extraction of the underlying signal a non-trivial statistical problem. Our inferential framework is {\em doubly-online} in the following sense. {First, we identify changepoints in a \emph{between-online} setting, in which activities are processed sequentially when a new one is fully observed. This permits to divide activities into subsequent segments and update the information on the unknown parameters at the end of each  activity. We also consider a {\em within-online} setting,  which refers to the online data processing of one activity. During a run, having information on the behavior difference between the current and the previous activities may be translated into motivational feedback or a potential alert before the end of the activity. Figure \ref{fig:sketch} shows the within-online setting for data collected by one runner in our dataset.  The red lines are associated to one new activity,  monitored by the athlete after five minutes of running and characterized by high effort, although the speed behavior seems to be similar to those in the previous activities (shown in gray). Our algorithm provides an online probabilistic quantification of the changepoint uncertainty by delivering the posterior probability of a behavioral change occurrence at any time point of the activity. In the case of Figure \ref{fig:sketch}, the runner is interested in the behavior change at minute $5$ of the current activity.
} 

%\subsection{Contribution, literature review and  structure of the paper}
We model the set of observed activities as a multivariate state space model \citep[][]{durbinkoopman12, shumway_book} and we adapt to this framework
classical changepoint modeling, which allows for the online detection of an a priori unknown set of changepoints between activities, see  \citet{chib98,fearnheadliu07,caron2012,Yildirim2013}. 
Changepoint detection is a relevant problem in many fields of science, ranging from industrial process control, health monitoring, cybersecurity, and machine learning \citep[see, e.g., ][]{Aminikhanghahi2017survey, titsias2020sequential, xie2021sequential, kaylea_fitbit}. 
Our approach differs in that we solve a problem of changepoint signal extraction in which the double sequential nature---between and within activities---of the data-generating process is preserved. The key idea is that we leverage the data on the past history of the athlete as a benchmark for identifying standard behaviors and deviations, providing relevant information about the performance as new data are collected.
In our application, making online inferences on a sequence of activities before the last one is fully observed is clearly of paramount importance. The literature on the changepoint detection problem is very large, and alternative approaches have been proposed for high dimensional frameworks, mostly based on dimensionality reduction techniques \citep[see, e.g.,][]{same_segm, grundy2020}. Such approaches, although potentially usable in the between-online setting, in which the observations for identifying changepoints consist of entire activities represented by multiple multivariate time series, are not directly applicable in the within-online setting, in which there is the need to preserve the dual sequential nature of the data. 
We contribute to this  literature by proposing a new state-space-based algorithm for changepoint detection in a sequence of time series by adopting the online Expectation-Maximization (EM) algorithm  developed by \cite{Yildirim2013}.  The nature of our problem requires taking into account three sources of dependence: one that inherits the sequential nature of subsequent activities, one that considers the autocorrelation structure within each activity, and  one that models the contemporaneous dependence between variables. As a byproduct of our model assumptions and the online inferential procedure, our approach processes sequences of data in a doubly-online framework. While classical changepoint models detect distributional changes in a sequence of activities (i.e., multivariate time series), our state space model coupled with the online EM approach provides the additional benefit of estimating the probability that a single activity is a changepoint during a run. 

The paper is organized as follows. Sections \ref{sec:model} and \ref{sec:algo}  describe the statistical model and the algorithm to perform inference respectively.  Section \ref{sec:sim} illustrates a small simulation study that assesses our algorithm performance with synthetic examples and Section \ref{sec:real} presents the results of our model using real data collected by an athlete. We conclude with Section \ref{sec:conclusion}, which describes the model's limitations and possbile future developments. Proofs and details on the available dataset are provided in the supplementary material. 
\section{The model}
\label{sec:model}

For each runner, we observe the data $\mathbf{y}_{1:N,1:T}$, composed of $N$ ordered activities that are represented by $P$-dimensional time series at $T$ time points.  An activity can be thought of as a running session taking place on different days; $T$ defines the duration of each activity, which is considered , for simplicity, to be equal for all activities, and $P$ denotes the number of smart device measurements, such as heart rate and speed. Our interest lies in modeling the data online and identifying changepoints during each activity, using information on both previous activities and previous recordings during the current activity.   We build our model by first introducing an $N$-dimensional latent vector $S_{1:N}=(S_1, \ldots, S_N)$ such that $S_1=1$
and $S_n-S_{n-1}=1$ if a changepoint occurs at the $n$-th ($n>2$) activity. The vector $S_{1:N}=(S_1, \ldots, S_N)$ divides the activities into $S_N$ contiguous \emph{segments}, in which activities belonging to different segments are assumed to be independent of each other.
%\added[id = MS]{Good for me the previous sentence, but I would change it a bit and write as follow:   Identification of these changepoints is achieved by introducing an $N$-dimensional latent vector $S_{1:N}=(S_1, \ldots, S_N)$, such that $S_1=1$
%and $S_n-S_{n-1}=1$ when a changepoint has occurred at the $n$-th ($n>2$) activity. The vector $S_{1:N}=(S_1, \ldots, S_N)$ divides the activities into $S_N$ contiguous \emph{segments}, where activities belonging to different segments are assumed to be independent of each other.}
The segments $S_{1:N}$ are modeled using a discrete state space Markov chain with transition probability
$p(S_n  \vert S_{n-1}) =  \lambda $
if $S_n =S_{n-1}+1$, for  $0<\lambda<1$. 

Assume that the activity $n$ belongs to segment $s$. We model its measurements at time $t$ by a state space representation with measurement equation 
\begin{equation}
\label{measurment_equation}
\mathbf{y}_{n,t} = \begin{bmatrix}\mathbf{Z}^{(S)}_{\boldsymbol{\theta}} & \mathbf{Z}^{(A)}_{\boldsymbol{\theta}}\end{bmatrix}\begin{bmatrix}
\boldsymbol{\alpha}^{(s)}_{t}\\
\boldsymbol{\alpha}_{n,t} 
\end{bmatrix} +\boldsymbol{\epsilon}_{n,t}, \end{equation}
with $\boldsymbol{\epsilon}_{n,t} \stackrel{iid}{\sim} N_P(\mathbf{0}, \mathbf{\Sigma}_{\boldsymbol{\theta}})$, and state equation
\begin{equation}
\label{state_equation}
\begin{bmatrix}
\boldsymbol{\alpha}_{t+1}^{(s)}\\
\boldsymbol{\alpha}_{n, t+1} 
\end{bmatrix}   = \begin{bmatrix}
\mathbf{T}^{(S)}_{\boldsymbol{\theta}} & \mathbf{0} \\\mathbf{0} & \mathbf{T}^{(A)}_{\boldsymbol{\theta}} \end{bmatrix}\begin{bmatrix}
\boldsymbol{\alpha}^{(s)}_{t}\\
\boldsymbol{\alpha}_{n,t} 
\end{bmatrix} +
\begin{bmatrix} \boldsymbol{\eta}^{(s)}_{t}\\ \boldsymbol{\eta}_{n,t} \end{bmatrix},
\end{equation}
with  $\boldsymbol{\eta}^{(s)}_t \stackrel{iid}{\sim} N_{M}(\mathbf{0}, \mathbf{\Psi}_{\boldsymbol{\theta}})$, $\boldsymbol{\eta}_{n, t} \stackrel{iid}{\sim} N_{K}(\mathbf{0}, \mathbf{\Delta}_{\boldsymbol{\theta}})$, and $\boldsymbol{\alpha}_{1}^{(s)} \stackrel{iid}{\sim} N_{M}(\hat{\boldsymbol{\alpha}}_{1\mid 0}^{(S)}, \mathbf{P}_{1\mid0}^{(S)} )$ independent of $\boldsymbol{\alpha}_{n,1} \stackrel{iid}{\sim} N_{K}(\hat{\boldsymbol{\alpha}}_{1\mid 0}^{(A)}, \mathbf{P}_{1\mid0}^{(A)})$. 
The subscript $\boldsymbol{\theta}$ is used throughout to highlight which parts of the model depend on, or are a function of, an unknown parameter vector $\boldsymbol{\theta} \in \mathbf{\Theta}$, which is the object of inference in the model.
The elements $\mathbf{Z}^{(S)}_{\boldsymbol{\theta}} $, $\mathbf{Z}^{(A)}_{\boldsymbol{\theta}},$ $\mathbf{T}^{(S)}_{\boldsymbol{\theta}}$, and $\mathbf{T}^{(A)}_{\boldsymbol{\theta}}$ are non-stochastic design matrices with  dimensions $P \times M$, $P \times  K$, $M \times M$, and $K \times K$, respectively. These matrices are shared across different segments and different activities, and may depend on  $\boldsymbol{\theta}$. Their specification is left undefined and depends on the specific application and behavior of the variables being considered, as it is typical in state space modeling \citep[see, e.g.,][]{durbinkoopman12}. Coupled with the design matrices, the covariance matrices  $\mathbf{\Sigma}_{\boldsymbol{\theta}}$, $\mathbf{\Psi}_{\boldsymbol{\theta}}$, and $\mathbf{\Delta}_{\boldsymbol{\theta}}$ of dimensions $P \times P$, $M \times M$, and $K \times K$, respectively, capture any contemporaneous dependencies between different elements of the model, such as the entries of the error component $\boldsymbol{\epsilon}_{n,t}$ of dimensions $P \times 1$  or those of the disturbance vectors $\boldsymbol{\eta}^{(s)}_{t}$ and $\boldsymbol{\eta}_{n,t}$  of dimensions $M \times 1$ and $K \times 1$, respectively. In general, the covariance matrices are full and unstructured; however, depending on the application, they may have a specific structure and involve a small number of elements of $\boldsymbol{\theta}$.

In the above specification, $\boldsymbol{\alpha}_t^{(s)}$ are vectors of dimensions $M \times 1$ that denote the dynamic segment-specific latent features, which are supposed to be independent of any other $\boldsymbol{\alpha}_t^{(s^\prime)}$, for any $s \neq s^\prime$. Together with $S_{1:N}$, the segment-specific latent features $\boldsymbol{\alpha}_t^{(s)}$ account for the dependence between subsequent activities. The activity-specific latent features $\boldsymbol{\alpha}_{n,t}$ are vectors of dimension $K \times 1$ that capture temporal dependencies that are unrelated to the performance of the athlete and describe negligible factors or disturbing aspects associated with the activities. These vectors are assumed to be independent of $\boldsymbol{\alpha}_t^{(s)}$ and any other $\boldsymbol{\alpha}_{n^\prime,t}$, with $n^\prime \neq n$. 
With no information on the initial states, we adopt the diffuse state initialization technique, in which the means and variances are independent  of $\boldsymbol{\theta}$, and the latter are supposed to be large \citep{durbinkoopman12}.

{Condition now on $S_{1:N}$ and assume further that the $s$-th segment ranges between the $j_s$-th and the $k_{s}$-th activity, so that its length is $m_s = k_s-j_s+1$.  We model this segment using the following equations:
\begin{align}
\label{s_model_meas}
    \begin{bmatrix}
    \mathbf{y}_{j_s,t}\\
    \mathbf{y}_{j_s+1,t}\\
    \vdots \\
    \mathbf{y}_{k_s,t}\\
    \end{bmatrix} = \begin{bmatrix}\mathbf{Z}^{(S)}_{\boldsymbol{\theta}} & \mathbf{Z}^{(A)}_{\boldsymbol{\theta}} & \mathbf{0} & \ldots & \mathbf{0}
    \\
    \mathbf{Z}^{(S)}_{\boldsymbol{\theta}} &  \mathbf{0} & \mathbf{Z}^{(A)}_{\boldsymbol{\theta}} & \mathbf{0} & \vdots \\
    \vdots & \vdots & \mathbf{0} & \ddots & \mathbf{0}
    \\
    \mathbf{Z}^{(S)}_{\boldsymbol{\theta}} &  \mathbf{0} & \ldots & \mathbf{0} & \mathbf{Z}^{(A)}_{\boldsymbol{\theta}} \\
    \end{bmatrix} \begin{bmatrix}
    \boldsymbol{\alpha}_t^{(s)}\\
    \boldsymbol{\alpha}_{j_s, t}\\
    \boldsymbol{\alpha}_{j_s+1,t}\\
 \vdots\\
    \boldsymbol{\alpha}_{k_s,t}\\
    \end{bmatrix} + \begin{bmatrix}\boldsymbol{\epsilon}_{j_s,t} \\ 
    \boldsymbol{\epsilon}_{j_s+1,t}\\
    \vdots\\
    \boldsymbol{\epsilon}_{k_s,t}
    \end{bmatrix},\end{align}
\begin{align}
\label{s_model_state}
\begin{bmatrix}
    \boldsymbol{\alpha}_{t+1}^{(s)}\\
    \boldsymbol{\alpha}_{j_s,t+1}\\
    \boldsymbol{\alpha}_{j_s+1,t+1}\\
 \vdots\\
    \boldsymbol{\alpha}_{k_s,t+1}\\
    \end{bmatrix} = \begin{bmatrix}\mathbf{T}^{(S)}_{\boldsymbol{\theta}} & \mathbf{0} & \ldots & \ldots &\mathbf{0}
    \\
    \mathbf{0} & \mathbf{T}^{(A)}_{\boldsymbol{\theta}}  & \mathbf{0} & \ldots & \mathbf{0} 
        \\
    \mathbf{0} &  \mathbf{0} & \mathbf{T}^{(A)}_{\boldsymbol{\theta}}  &  \mathbf{0}& \mathbf{0} 
    \\
    \mathbf{0} & \mathbf{0} & \mathbf{0}  &  \ddots & \mathbf{0} 
    \\
    \mathbf{0} & \mathbf{0} & \ldots  & \mathbf{0}  & \mathbf{T}^{(A)}_{\boldsymbol{\theta}}  
    \end{bmatrix}     \begin{bmatrix}
    \boldsymbol{\alpha}_{t}^{(s)}\\
    \boldsymbol{\alpha}_{j_s,t}\\
    \boldsymbol{\alpha}_{j_s+1,t}\\
 \vdots\\
    \boldsymbol{\alpha}_{k_s,t}\\
    \end{bmatrix}  + \begin{bmatrix}
    \boldsymbol{\eta}_{t}^{(s)}\\
    \boldsymbol{\eta}_{j_s,t}\\
    \boldsymbol{\eta}_{j_s+1,t}\\
 \vdots\\
    \boldsymbol{\eta}_{k_s,t}\\
    \end{bmatrix}
\end{align}
for $\boldsymbol{\epsilon}_{j_s:k_s,t} = (\boldsymbol{\epsilon}_{j_s,t}^\prime, \boldsymbol{\epsilon}_{j_s+1,t}^\prime, \ldots, \boldsymbol{\epsilon}_{k_s,t}^\prime)^\prime \stackrel{iid}{\sim} N_{m_s P}(\mathbf{0}, \mathbf{I}_{m_s} \otimes \mathbf{\Sigma}_{\boldsymbol{\theta}})$, 
$\boldsymbol{\eta}_{j_s:k_s,t} = (\boldsymbol{\eta}_{j_s,t}^\prime,\boldsymbol{\eta}_{j_s+1,t}^\prime \ldots,\boldsymbol{\eta}_{k_s,t}^\prime)^\prime \stackrel{iid}{\sim} N_{m_s K}( \mathbf{0},  \mathbf{I}_{m_s} \otimes \mathbf{\Delta}_{\boldsymbol{\theta}})$, 
$\eta_t^{(s)} \sim N_{M}(\mathbf{0}, \mathbf{\Psi}_{\boldsymbol{\theta}})$, and $\boldsymbol{\alpha}_{j_s:k_s,1} = (\boldsymbol{\alpha}_{j_s,1}^\prime,\boldsymbol{\alpha}_{j_s+1,1}^\prime \ldots,\boldsymbol{\alpha}_{k_s,1}^\prime)^\prime {\sim} N_{m_s K}( \mathbf{1}_{m_s} \otimes \hat{\boldsymbol{\alpha}}_{1 \vert 0}^{(A)},  \mathbf{I}_{m_s} \otimes \mathbf{P}_{1 \vert 0}^{(A)})$, $\boldsymbol{\alpha}_{1}^{(s)} \sim N_{M}(\hat{\boldsymbol{\alpha}}_{1 \vert 0}^{(S)}, \mathbf{P}_{1 \vert 0}^{(S)})$, independent of each other and with fixed hyper-parameters.}

{Let $\boldsymbol{\alpha}_{1:T}^{1:N}= (\boldsymbol{\alpha}_{1:N,1:T}, \boldsymbol{\alpha}_{1:T}^{(1:S_N)})$ be a vector storing both the segment-specific and the activity-specific latent features.  It is possible to write the \emph{augmented} likelihood of the model, which has the conditional independence structure
\begin{align}
\label{augmented}
p_{\boldsymbol{\theta}}(\mathbf{y}_{1:N,1:T}, \boldsymbol{\alpha}_{1:T}^{1:N}, S_{1:N}) =  p_{\boldsymbol{\theta}}(\mathbf{y}_{1:N,1:T} \vert \boldsymbol{\alpha}_{1:T}^{1:N},  S_{1:N}) p_{\boldsymbol{\theta}}(\boldsymbol{\alpha}_{1:T}^{1:N} \vert S_{1:N})p(S_{1:N}),
\end{align}
where $p_{\boldsymbol{\theta}}(\boldsymbol{\alpha}_{1:T}^{1:N} \vert S_{1:N}) = p_{\boldsymbol{\theta}}(\boldsymbol{\alpha}_{1:T}^{(1:S_N)} \vert S_{1:N}) p_{\boldsymbol{\theta}}(\boldsymbol{\alpha}_{1:N,1:T})$.}
Conditional on segments $S_{1:N}$, Equations \eqref{s_model_meas} and \eqref{s_model_state} specify a state space model such that both the segment-specific and activity-specific latent features can be integrated out by means of a Kalman filter routine. By integrating out these latent features in Equation \eqref{augmented} we  obtain the contribution of the $s$-th segment to the likelihood conditional on $S_{1:N}$ given by
\begin{align}
\label{m;cond_lik}
    \log p_{\boldsymbol{\theta}}(\mathbf{y}_{j_s:k_s,1:T} \vert S_{j_s:k_s}) = 
    -\frac{1}{2} \sum_{t = 1}^T   \bigg(m_s P  \log(2\pi) + \log | \mathbf{F}_{s,t} | + \boldsymbol{\upsilon}_{j_s:k_s,t}^\prime (\mathbf{F}_{s,t})^{-1} \boldsymbol{\upsilon}_{j_s:k_s,t}\bigg)
\end{align}
{where both the innovations vectors $\boldsymbol{\upsilon}_{j_s:k_s,t}$ and their respective covariance matrices $\mathbf{F}_{s,t}$ are outputs of the Kalman filter routine, reviewed in the supplementary material. Thus, the likelihood is conditional on the segments, but no longer on the segment- and activity-specific latent features. The conditional likelihood depends clearly on the unknown parameter $\boldsymbol{\theta}$ through $\boldsymbol{\upsilon}_{j_s:k_s,t}$  and $\mathbf{F}_{s,t}$, which are functions of the data, the design matrices, and the covariance matrices involved in the state space model, for which the subscript $\boldsymbol{\theta}$ has been omitted for simplicity of notation.} 
{While the model specification above is intuitively driven by the mechanism that generates the data, it is useful to connect it with the way  \cite{Yildirim2013} specified a model because we will adopt their inferential strategy in the next section. Specifically, instead of $S_{1:N}$, we can define a latent vector $D_{1:N} = (D_1,\ldots,D_N)$ such that $D_n$ represents the delay of from the last changepoint defined through the following recursion
\begin{align*}
    D_n \vert D_{n-1} = \begin{cases} D_{n-1}+1 &\quad \text{if } \quad S_n = S_{n-1}\\ 1 &\quad \text{if } \quad S_n = S_{n-1}+1
    \end{cases},
\end{align*}
with $D_1 = 1$, and we note the information equivalence between $D_{1:N}$ and $S_{1:N}$. We can then express the conditional likelihood of the observed process as
\begin{align}
\label{cond_likelihood_potential}
p_{\boldsymbol{\theta}}(\mathbf{y}_{1:N,1:T} \vert D_{1:N}) =   \prod_{n = 1}^{N}  G_{\boldsymbol{\theta},n}^D(D_n),
\end{align}
where the \emph{potentials} are defined as
\begin{align*}
    G_{\boldsymbol{\theta},n}^D (D_n) =   p_{\boldsymbol{\theta}}(\mathbf{y}_{n,1:T} \vert D_{1:n},\mathbf{y}_{1:(n-1),1:T} )  = \begin{cases}\frac{p_{\boldsymbol{\theta}}(\mathbf{y}_{j:n,1:T} \vert D_n)}{p_{\boldsymbol{\theta}}(\mathbf{y}_{j:(n-1),1:T} \vert D_{n-1} )} \quad &\text{if } D_n = D_{n-1}+1 \\
    p_{\boldsymbol{\theta}}(\mathbf{y}_{n,1:T} \vert D_{n}) \quad  &\text{if } D_n = 1
    \end{cases},
\end{align*}
with $j = n-D_n +1$. Notice that the potential $G_{\boldsymbol{\theta},n}^D (D_n)$ is nothing more than the individual contribution of activity $n$ to the conditional likelihood of the observed process, provided that the first $n-1$ activities have already been observed and the index of the last changepoint is known by means of $D_n$.
The likelihoods involved in the potentials can be easily calculated through the use of Kalman filter routines, as in Equation \eqref{m;cond_lik}, in which, for activity $n$, the activities to be considered in the respective segment are determined by $D_n$. Knowing either $D_{1:N}$  or $S_{1:N}$ is equivalent, while if we consider only the marginal $D_n$ instead of $S_{j:n}$ with $j = \text{max}(1, S_n-D_n+1)$, we lose the information on the number of the segment the $n$-th activity belongs to. We do not consider the random variable $D_n^{s}$, which highlights both the delay with respect to the last changepoint and the segment to which the activity belongs to. Since our primary interest is the early changepoint detection, all the provided results rely on an underlying exchangeability assumption between segment-specific features, which simplifies the mathematical treatment.

The likelihood of the observed process is given by
%\begin{align*}
%\label{marginal_likelihood_delays}
    $p_{\boldsymbol{\theta}}(\mathbf{y}_{1:N,1:T}) = \text{E}_{\boldsymbol{\theta}} \big[ \prod_{n = 1}^{N}  G_{\boldsymbol{\theta},n}^D (D_n)\big]$
%\end{align*}
where the expectation is taken with respect to $D_{1:N}$. 
This likelihood represents the target to maximize for obtaining an estimate of the unknown parameter $\boldsymbol{\theta}$, which drives the behavior of the observed process. The parameter $\boldsymbol{\theta}$ is involved in the model specification of both the segments-specific, and the activity-specific temporal dynamics during the activities.

}
\section{Estimation and changepoint detection}
\label{sec:algo}
\subsection{From batch to online EM algorithms}
Our interest lies in 
$
\hat{\boldsymbol{\theta}} = \underset{\boldsymbol{\theta} \in \mathbf{\Theta}}{\text{arg max} }\big[ p_{\boldsymbol{\theta}}(\mathbf{y}_{1:N, 1:T})\big]
$
via the EM algorithm introduced by \cite{dempster77}.  An exact online EM algorithm for linear and Gaussian state space models was introduced by \cite{elliot2002}. Here, we review and adapt to our setting the online EM algorithm by \cite{Yildirim2013}, involving a Sequential Monte Carlo (SMC) approximation step, developed for a large class of changepoints models.

Let $\hat{\boldsymbol{\theta}}_{it}$ be the the estimate of the maximizer at the $it$-th iteration of the EM algorithm. At iteration $it+1$ the expectation step of the offline EM algorithm computes 
\begin{align}
\label{E_step1}
    Q_{1:N} & (\boldsymbol{\theta}, \hat{\boldsymbol{\theta}}_{it})  = \text{E}_{\hat{\boldsymbol{\theta}}_{it}}\big[ \log p_{\boldsymbol{\theta}}(\mathbf{y}_{1:N}, \boldsymbol{\alpha}^{1:N}_{ 1:T}, D_{1:N})\vert \mathbf{y}_{1:N, 1:T}\big] \\
 \label{E_step2}
& =  \text{E}_{\hat{\boldsymbol{\theta}}_{it}} \bigg[  \log p(D_{1:N}) + \text{E}_{\hat{\boldsymbol{\theta}}_{it}}\big[\log p_{\boldsymbol{\theta}}(\mathbf{y}_{1:N}, \boldsymbol{\alpha}^{1:N}_{ 1:T} \vert D_{1:N}) \vert D_{1:N}, \mathbf{y}_{1:N,1:T}\big] \vert \mathbf{y}_{1:N,1:T}  \bigg]
\end{align}
The expected value in Equation \eqref{E_step1} is computed with respect to both $D_{1:N}$ and the latent features $\boldsymbol{\alpha}^{1:N}_{1:T}$, considered jointly, and involves the log-density augmented for both latent variables. Equation \eqref{E_step2}  involves an external and an internal expectation, which are computed with respect to the random variables $D_{1:N}$ and $\boldsymbol{\alpha}_{1:T}^{1:N} \vert D_{1:N}$, respectively, given the entire set of data $\mathbf{y}_{1:N,1:T}$. 
The subscript $1$:$N$ in $Q_{1:N}(\boldsymbol{\theta}, \hat{\boldsymbol{\theta}}_{it})$ indicates that all the observations up to activity $N$ are used.  Moreover, $Q_{1:N}(\boldsymbol{\theta}, \hat{\boldsymbol{\theta}}_{it})$ depends on $\boldsymbol{\theta}$ through the functional form of the augmented likelihood $p_{\boldsymbol{\theta}}(\mathbf{y}_{1:N}, \boldsymbol{\alpha}^{1:N}_{ 1:T}, D_{1:N})$. The true parameter $\boldsymbol{\theta}$ is substituted by its estimate $\hat{\boldsymbol{\theta}}_{it}$ when the expected values are computed at iteration $it+1$. 
Once this expectation is computed, the maximization step solves 
\begin{align}
\label{M_step1}
    \hat{\boldsymbol{\theta}}_{it+1} = \underset{\boldsymbol{\theta} \in \mathbf{\Theta}}{\text{arg max} } \big[Q_{1:N}(\boldsymbol{\theta}, \hat{\boldsymbol{\theta}}_{it})\big]  = \mathbf{\Lambda}(\mathcal{Q}_{1:N})
\end{align}
with $\mathbf{\Lambda}:\mathcal{Q}_{1:N} \rightarrow{\mathbf{\Theta}}$, and $\mathcal{Q}_{1:N}$ being the $r$-dimensional set of sufficient statistics. 
The two steps are repeated until a set of stopping rules are satisfied, which allows to iteratively grow the function $Q_{1:N}(\boldsymbol{\theta}, \hat{\boldsymbol{\theta}}_{it})$ and, consequently, the likelihood of the observed process. The offline EM algorithm requires the ability to compute both the E-step in Equation \eqref{E_step1} 
and the M-step in Equation \eqref{M_step1} 
in closed form or through the use of a finite set of elementary operations, involving the expectation of the set of $r$ sufficient statistics $\mathcal{Q}_{1:N}$.

To adapt the EM algorithm to the online setting, we define the {\emph{individual contribution}} of activity $n$ to $Q_{1:n}(\boldsymbol{\theta},\boldsymbol{\theta}^\prime)$   as
\begin{align*}
    \iota_{\boldsymbol{\theta}^\prime}(\mathbf{y}_{n,1:T}) & := \log p(D_{1:n}) - \log p(D_{1:(n-1)}) \nonumber \\ &\quad + \text{E}_{\boldsymbol{\theta}^\prime}\big[ \log p_{\boldsymbol{\theta}}(\mathbf{y}_{1:n,1:T}, \boldsymbol{\alpha}^{1:n}_{ 1:T} \vert D_{1:n}) \vert \mathbf{y}_{1:n, 1:T}, D_{1:n}\big]  \\& \quad  -\text{E}_{\boldsymbol{\theta}^\prime}\big[ \log p_{\boldsymbol{\theta}}(\mathbf{y}_{1:(n-1),1:T}, \boldsymbol{\alpha}^{1:(n-1)}_{ 1:T}\vert  D_{1:(n-1)}) \vert \mathbf{y}_{1:(n-1), 1:T},  D_{1:(n-1) }\big],
\end{align*}
{with $\iota_{\boldsymbol{\theta}^\prime}(\mathbf{y}_{1,1:T}) = \text{I}(D_1 = 1) + \text{E}_{\boldsymbol{\theta}^\prime}\big[ \log p_{\boldsymbol{\theta}}(\mathbf{y}_{1,1:T}, \boldsymbol{\alpha}^{1}_{1:T}\vert D_{1})\vert \mathbf{y}_{1,1:T}, D_{1} \big]$, for any value $\boldsymbol{\theta}^\prime \in \mathbf{\Theta}$. The expression for $\iota_{\boldsymbol{\theta}^\prime}(\mathbf{y}_{n,1:T})$ is nothing else but the difference between the argument of the external expected value in Equation \eqref{E_step2} computed using the observations up to activity $n$ and the same argument calculated using with observations up to activity $n-1$, in which the expectations are taken with respect to the latent features $\boldsymbol{\alpha}^{1:n}_{ 1:T}$ and $\boldsymbol{\alpha}^{1:(n-1)}_{ 1:T}$ involved in the respective state space models.}
{Although not easy to interpret, the construction of $\iota_{\boldsymbol{\theta}^\prime}(\mathbf{y}_{n,1:T})$ mimics the definition of the conditional likelihood in terms of the potentials in Equation \eqref{cond_likelihood_potential} and allows to write the expression of   $Q_{1:N}(\boldsymbol{\theta},\boldsymbol{\theta}^\prime)$ as the expected value  with respect to $D_{1:N}$ of a sum of $N$ functionals, i.e.
$
Q_{1:N}(\boldsymbol{\theta},\boldsymbol{\theta}^\prime) = \text{E}_{\boldsymbol{\theta}^\prime} \big[\sum_{n = 1}^{N}  \iota_{\boldsymbol{\theta}^\prime}(\mathbf{y}_{n,1:T}) \vert \mathbf{y}_{1:N,1:T} \big]
$, and therefore its sequential evaluation  as new activities are observed.}

We adopt the stochastic approximation proposed by \cite{Yildirim2013} based on a forward smoothing technique, see for example \cite{kantas15}.  By setting $\mathbf{T}_1(D_1, \boldsymbol{\theta}) = \iota_{\boldsymbol{\theta}}(\mathbf{y}_{1,1:T})$, and defining
\begin{align}
    %\label{smc_S1}
    \mathbf{S}_n(D_{1:n},   \boldsymbol{\theta}^\prime) : & = \sum_{j = 1}^{n} \iota_{\boldsymbol{\theta}^\prime }(\mathbf{y}_{j,1:T}) \nonumber
    \\
    %\label{smc_T1}
    \mathbf{T}_n(D_{1:n}, \boldsymbol{\theta}^\prime) : &= \sum_{D_{1:(n-1)} \in \mathcal{D}_{1:(n-1)}} \mathbf{S}_n(D_{1:n},   \boldsymbol{\theta}^\prime ) p_{\boldsymbol{\theta}}(D_{1:(n-1)}\vert \mathbf{y}_{1:(n-1),1:T}, D_{n}) \nonumber \\
    \label{smc_T2}
    & = \sum_{D_{n-1} \in \mathcal{D}_{n-1}} \big[  \mathbf{T}_{n-1}(D_{1:(n-1)}, \boldsymbol{\theta}^\prime) + \iota_{\boldsymbol{\theta}^\prime}(\mathbf{y}_{n,1:T}) \big] p_{\boldsymbol{\theta}}(D_{n-1} \vert \mathbf{y}_{1:(n-1),1:T}, D_n),
\end{align}
we are able to evaluate $\mathbf{T}_n(D_{1:n}, \boldsymbol{\theta}^\prime)$ sequentially. It can also be shown that
\begin{align*}
%\label{T_eval}
{Q_{1:n}(\boldsymbol{\theta},\boldsymbol{\theta}^\prime)=} \text{E}_{\boldsymbol{\theta}^\prime} \big[\sum_{ j= 1}^{n} \iota_{\boldsymbol{\theta}^\prime}(\mathbf{y}_{j,1:T}) \vert \mathbf{y}_{1:n,1:T} \big] = \sum_{D_n \in \mathcal{D}_{n}} \mathbf{T}_n(D_{1:n}, \boldsymbol{\theta}^\prime) p_{\boldsymbol{\theta}}(D_n \vert \mathbf{y}_{1:n, 1:T})
\end{align*}
{allowing, subject to knowing $\mathbf{T}_n(D_{1:n}, \boldsymbol{\theta}^\prime)$ and $p_{\boldsymbol{\theta}}(D_n \vert \mathbf{y}_{1:n, 1:T})$, to also obtain $Q_{1:n}(\boldsymbol{\theta},\boldsymbol{\theta}^\prime)$ sequentially for any activity $n$ that has been fully observed.}

Let $\gamma_n$ be a step-size decreasing function such that $0<\gamma_n<1$, $\sum_{n = 1}^{\infty} \gamma_n = \infty$,  $\sum_{n = 1}^{\infty} \gamma_n^2 < \infty$. The stochastic approximation of Equation \eqref{smc_T2}  proposed by \cite{Yildirim2013} becomes 
\begin{align}
    \label{T_approx}
    \mathbf{T}_{\gamma, n}(D_{1:n};\hat{\boldsymbol{\theta}}_{n-1} ) = \sum_{D_{n-1} \in \mathcal{D}_{n-1}} &\big[(1-\gamma_n)     \mathbf{T}_{\gamma, n-1}(D_{1:(n-1)}; \hat{\boldsymbol{\theta}}_{n-2})\nonumber  \\ & + \gamma_n \iota_{\hat{\boldsymbol{\theta}}_{n-1}}(\mathbf{y}_{n,1:T})\big]    p_{\hat{\boldsymbol{\theta}}_{1:(n-1)}}(D_{n-1} \vert \mathbf{y}_{1:(n-1),1:T}, D_n),
\end{align}
which leads to 
\begin{align*}
    %\label{expect_approx}
    \mathcal{Q}_n  = \sum_{D_n \in \mathcal{D}_n}       \mathbf{T}_{\gamma, n}(D_{1:n};\hat{\boldsymbol{\theta}}_{n-1} ) p_{\hat{\boldsymbol{\theta}}_{1:(n-1)}}(D_n \vert \mathbf{y}_{1:n, 1:T}),
\end{align*}
which is used for obtaining $\hat{\boldsymbol{\theta}}_{n}$, in substitution of $\mathcal{Q}_{1:N}$ in Equation\eqref{M_step1}. The algorithm requires the ability to compute online the approximations $p_{\hat{\boldsymbol{\theta}}_{1:(n-1)}}(D_{n-1} \vert \mathbf{y}_{1:(n-1)}, D_n)$ and  $p_{\hat{\boldsymbol{\theta}}_{1:(n-1)}}(D_n \vert \mathbf{y}_{1:(n-1), 1:T})$, obtained here by an SMC approximation,
as described in the next subsection.

\subsection{SMC approximation of the predicted probabilities}
{The between-online setting processes activity the data sequentially, whenever an activity has been fully observed. The purpose of the between-online setting is to leverage existing proposals in the literature for online parameter estimation and changepoint identification, which is useful both for retrospective performance analysis and as an analysis toold in the within-online setting.} We review here the principles underlying the algorithm proposed by \cite{Yildirim2013} {and derive the computations that lead to our algorithm for changepoint detection, details of which are given in the supplementary material.}
Suppose that $p_{\boldsymbol{\theta}}(D_{n-1}, \vert \mathbf{y}_{1:(n-1),1:T})$ is known. The quantity
\begin{align}
%\label{predi1}
    p_{\boldsymbol{\theta}}(D_{n}, \vert \mathbf{y}_{1:(n-1),1:T}) &= \sum_{D_{n-1} \in \mathcal{D}_{n-1}} p_{\boldsymbol{\theta}}(D_{n}, D_{n-1} \vert \mathbf{y}_{1:(n-1),1:T}) \nonumber\\
\label{predi2}
& =\sum_{D_{n-1} \in \mathcal{D}_{n-1}}p(D_{n} \vert D_{n-1})  p_{\boldsymbol{\theta}}(D_{n-1} \vert \mathbf{y}_{1:(n-1),1:T})
%\\
% & = \sum_{D_{n-1} \in \mathcal{D}^{n-1}}p_{\boldsymbol{\theta}}(D_n \vert D_{n-1} ) G_{\boldsymbol{\theta},n} (D_{n-1} )  p_{\boldsymbol{\theta}}(D_{n-1} \vert \mathbf{y}_{1:(n-2), 1:T}) \nonumber
\end{align}
can be used to derive exactly
\begin{align*}
%\label{smo_prob}
     p_{\boldsymbol{\theta}}( D_{n-1} \vert D_n, \mathbf{y}_{1:(n-1),1:T})  & = \frac{p_{\boldsymbol{\theta}}(D_{n}, D_{n-1} \vert \mathbf{y}_{1:(n-1),1:T})}{\sum_{D_{n-1}^\prime \in \mathcal{D}_{n-1}}p_{\boldsymbol{\theta}}(D_{n}, D_{n-1}^\prime \vert \mathbf{y}_{1:(n-1),1:T})}%\\
    % & =  \frac{p(D_{n} \vert  D_{n-1})  p_{\boldsymbol{\theta}}(D_{n-1} \vert \mathbf{y}_{1:(n-1),1:T})}{\sum_{D_{n-1}^\prime \in \mathcal{D}_{n-1}} p(D_{n} \vert  D_{n-1}^\prime) p_{\boldsymbol{\theta}}(D_{n-1} \vert \mathbf{y}_{1:(n-1),1:T})} 
 \\ &= \frac{p_{\boldsymbol{\theta}}(D_n \vert D_{n-1}) G_{\boldsymbol{\theta},n} (D_{n-1})  p_{\boldsymbol{\theta}}(D_{n-1} \vert \mathbf{y}_{1:(n-2), 1:T})}{\sum_{D_{n-1}^\prime \in \mathcal{D}^{n-1}}p_{\boldsymbol{\theta}}(D_n \vert D_{n-1}^\prime ) G_{\boldsymbol{\theta},n} (D_{n-1}^\prime )  p_{\boldsymbol{\theta}}(D_{n-1}^\prime \vert \mathbf{y}_{1:(n-2), 1:T})}
\end{align*}
and
\begin{align*}
%\label{filt_prob}
 p_{\boldsymbol{\theta}}(D_{n}\vert \mathbf{y}_{1:n,1:T}) & = \frac{G_{\boldsymbol{\theta}, n}^D(D_n) p_{\boldsymbol{\theta}}(D_{n} \vert \mathbf{y}_{1:(n-1),1:T})}{\sum_{D_{n}^\prime \in \mathcal{D}_{n}}G_{\boldsymbol{\theta}, n}^D(D_n) p_{\boldsymbol{\theta}}(D_{n}^\prime,  \vert \mathbf{y}_{1:(n-1),1:T})}
\end{align*}
where $G_{\boldsymbol{\theta}, n}^D(D_n) = p_{\boldsymbol{\theta}}(\mathbf{y}_{n,1:T} \vert D_{n}, \mathbf{y}_{1:(n-1),1:T}) $.
It is important to note that, although the involved quantities can be obtained exactly, computing Equation \eqref{predi2} has complexity $O(n)$, as $p(D_{n} \vert D_{n-1})  \neq 0$ for $2(n-1)$ combinations of $(D_{n} , D_{n-1})$. Hence, the online exact computation for a large panel of activities may be impractical in many situations, as the complexity increases with new activities.

 Let $\eta_{n-1}^B(D_{n-1})$ be a particle approximation of $p_{\boldsymbol{\theta}}(D_{n-1} \vert \mathbf{y}_{1:(n-2),1:T})$, composed of $B$ particles with support $\mathcal{D}_{n-1}^B = \{d_{n-1}^1, \ldots, d_{n-1}^B, d_{n-1}^B\}$ composed by the particles themselves. Consider then the \emph{augmented support} $\mathcal{D}_n^{B\star}$ of dimension $2B$ defined as
\begin{align*}
    \mathcal{D}_n^{B\star} = \{ (1, d_{n-1}^1), (d_{n-1}^1+ 1, d_{n-1}^1), \ldots,   (1, d_{n-1}^B), (d_{n-1}^B+ 1, d_{n-1}^B)\}.
\end{align*}
An approximation of $p_{\boldsymbol{\theta}}(D_{n} \vert \mathbf{y}_{1:(n-1),1:T})$ can be obtained by sampling $B$ independent particles from $\mathcal{D}_n^{B\star}$ with weight
$
W(D_{n},D_{n-1}) \propto p(D_{n} \vert D_{n-1}) G_{\boldsymbol{\theta},n-1}^D(D_{n-1}) \eta_{n-1}^B(D_{n-1})$, 
and then marginalizing with respect to $D_{n-1}$. Let $\mathcal{D}_{(n, n-1)}^B = \{(d_{n}^1, d_{n-1}^1), \ldots, (d_{n}^B,d_{n-1}^B)\}$ be the $B$ sampled particles. The approximation of $p_{\boldsymbol{\theta}}(D_{n} \vert \mathbf{y}_{1:(n-1),1:T})$ is
$
   \eta_{n}^B(D_{n}) = \sum_{b = 1}^B \delta_{D_n}(d_{n}^b, d_{n-1}^b),
$
with support $\mathcal{D}_n^{B} = \{d_{n}^1 \ldots, d_{n}^B\}$, where $ \delta_{D_n}(d_{n}^b, d_{n-1}^b) = 1$ if $D_n = d_n^b$, and $0$ otherwise.
Moreover, $p_{\boldsymbol{\theta}}( D_{n-1} \vert D_n, \mathbf{y}_{1:(n-1),1:T})$ is approximated by
\begin{align*}
p_{\hat{\boldsymbol{\theta}}_{1:(n-1)}}(D_{n-1} \vert \mathbf{y}_{1:(n-1),1:T}, D_n) = \frac{p(D_{n} \vert  D_{n-1})G_{\hat{\boldsymbol{\theta}}_{n-1}, n-1}^{D}(D_{n-1})   \eta_{n-1}^B(D_{n-1})}{\sum_{D_{n-1}^\prime \in \mathcal{D}_{n-1 \vert n}^B}p(D_{n} \vert  D_{n-1})G_{\hat{\boldsymbol{\theta}}_{n-1}, n-1}^{D}(D_{n-1}^\prime) \eta_{n-1}^B(D_{n-1})},
\end{align*}
and $ p_{{\boldsymbol{\theta}} }(D_n \vert \mathbf{y}_{1:n, 1:T})$ by
\begin{align}
\label{filtered approx}
    p_{\hat{\boldsymbol{\theta}}_{1:(n-1)}}(D_n \vert \mathbf{y}_{1:n, 1:T}) = \frac{\sum_{D_{n-1} \in \mathcal{D}_{n-1}^B} G_{\hat{\boldsymbol{\theta}}_{n-1}, n}^{D}(D_n) p(D_{n} \vert  D_{n-1}) \eta_{n-1}^B(D_{n-1}) }{\sum_{(D_{n}^\prime, D_{n-1}^\prime) \in \mathcal{D}_{(n, n-1)}^B} G_{\hat{\boldsymbol{\theta}}_{n-1}, n}^{D}(D_n^\prime) p(D_{n}^\prime \vert  D_{n-1}^\prime) \eta_{n-1}^B(D_{n-1}^\prime)},
\end{align}
over the supports $\mathcal{D}_{(n, n-1)}^B$ and $\mathcal{D}_{n}^B$, respectively, where the index $\hat{\boldsymbol{\theta}}_{1:(n-1)}$ highlights the fact that the approximations are obtained via a sequence of parameter's updates.

\subsubsection{Maximization step and inner expectations}
The maximization step in Equation \eqref{M_step1} that attempts to solve
$
    \frac{\partial     Q_{1:N}(\boldsymbol{\theta}, \boldsymbol{\theta}^\prime)}{ \partial \boldsymbol{\theta}}= \mathbf{0}
$
requires the computation of the derivative with respect to the elements  $\mathbf{Z}^{(S)}_{\boldsymbol{\theta}}$, $\mathbf{Z}^{(A)}_{\boldsymbol{\theta}}$, $\mathbf{T}^{(S)}_{\boldsymbol{\theta}}$, $\mathbf{T}^{(A)}_{\boldsymbol{\theta}}$, $\mathbf{\Delta}_{\boldsymbol{\theta}}$, $\mathbf{\Sigma}_{\boldsymbol{\theta}}$, and $\mathbf{\Psi}_{\boldsymbol{\theta}}$, before applying the chain rule to obtain the derivative with respect to ${\boldsymbol{\theta}}$. These computations involve a finite set of elementary operations and the knowledge of both the inner and the outer expectations in Equation \eqref{E_step2}. In the online setting, the SMC approximation allows to compute the outer expectation conditional on the available data, while the inner expectation can be obtained by considering that, conditioned on $S_{1:n}$, the model for the $s$--th segment in Equations \eqref{s_model_meas} and \eqref{s_model_state} is a linear Gaussian state space model.

These quantities can be generally obtained by standard  Kalman recursions, such as the Kalman smoother and the lagged smoother proposed, for example, by \cite{durbinkoopman12} and \cite{shumway_book}. Indeed, let us condition on $S_{1:n}$ or, equivalently, on the sequence of delays $D_{1:n}$. By the independence assumption between activities of different segments, it can be shown that  $\iota_{\boldsymbol{\theta}}(\mathbf{y}_{n,1:T})$ depends only on the activities that belong to the last segment. Let us define $L_{\boldsymbol{\theta}^\prime}(\mathbf{y}_{j:n,1:T}) = \text{E}_{\boldsymbol{\theta}^\prime}\big[ \log p_{\boldsymbol{\theta}}(\mathbf{y}_{j:n,1:T},\boldsymbol{\alpha}^{{j:n}}_{1:T} \vert D_{n}) \vert \mathbf{y}_{j:n, 1:T}, D_{n}\big]$, with $j = \max(1, n-D_n+1)$. The quantity $\iota_{\boldsymbol{\theta}^\prime}(\mathbf{y}_{n,1:T})$ is exactly
\begin{align*}
    \iota_{\boldsymbol{\theta}^\prime}(\mathbf{y}_{n,1:T}) =  \begin{cases} 1-\lambda+L_{\boldsymbol{\theta}^\prime}(\mathbf{y}_{j:n,1:T}) -L_{\boldsymbol{\theta}^\prime}(\mathbf{y}_{j:(n-1),1:T}) & \quad \text{if } j = \max(1, n-D_n+1)<n \\
    \lambda +L_{\boldsymbol{\theta}^\prime}(\mathbf{y}_{j:n,1:T}) &\quad \text{if }j = \max(1, n-D_n+1)=n,
    \end{cases} 
\end{align*}
where $L_{\boldsymbol{\theta}^\prime}(\mathbf{y}_{j:n,1:T})$ depends on the expectations
$\text{E}_{\boldsymbol{\theta}^\prime}\big[\boldsymbol{\alpha}_{t}^{j:n}\vert D_{n}, \mathbf{y}_{j:n,1:T} \big]$,      $\text{E}_{\boldsymbol{\theta}^\prime}\big[(\boldsymbol{\alpha}_{t}^{j:n})(\boldsymbol{\alpha}_{t}^{j:n})^\prime \vert D_{n}, \mathbf{y}_{j:n,1:T} \big]$, and
$\text{E}_{\boldsymbol{\theta}^\prime}\big[(\boldsymbol{\alpha}_{t+1}^{j:n})(\boldsymbol{\alpha}_{t}^{j:n})^\prime \vert D_{n}, \mathbf{y}_{j:n,1:T} \big]$,
which are computedusing the standard Kalman filtering, smoothing, and lagged smoothing routines, {reviewed in the supplementary material} \citep[see, e.g.,][]{shumway_book, durbinkoopman12}.

\subsection{Monitoring new activities in the within-online setting}

The ability to monitor the presence of a changepoint during activity $n$ is given by the need of computing, on the fly, $    p_{\boldsymbol{\theta}}(D_n\vert \mathbf{y}_{n,1:t}, \mathbf{y}_{1:(n-1),1:T})$
for any $t<T$. Note that  the activities $\mathbf{y}_{1:(n-1),1:T}$ have already been observed completely, $\mathbf{y}_{n,1:t}$ is the $n$-th activity that is being observed, and the interest resides in checking whether $D_n= 1$ or not.  This allows knowledge of the status of the athlete during an activity, while also accounting for their already observed past. The direct use of the Bayes formula gives 
\begin{align*}
p_{\boldsymbol{\theta}}(D_n \vert \mathbf{y}_{n,1:t}, \mathbf{y}_{1:(n-1),1:T}) = \frac{p_{\boldsymbol{\theta}}(\mathbf{y}_{n,1:t} \vert D_n, \mathbf{y}_{1:(n-1),1:T})p_{\boldsymbol{\theta}}(D_n \vert \mathbf{y}_{1:(n-1),1:T})}{\sum_{D_n^\prime \in \mathcal{D}^n}p_{\boldsymbol{\theta}}(\mathbf{y}_{n,1:t} \vert D_n^\prime, \mathbf{y}_{1:(n-1),1:T})p_{\boldsymbol{\theta}}(D_n^\prime \vert\mathbf{y}_{1:(n-1),1:T})}.
\end{align*}
An approximation of the predicted probability $p_{\boldsymbol{\theta}}(D_n \vert \mathbf{y}_{1:(n-1),1:T})$ is given by the SMC approach used by our algorithm so that we now consider the element $p_{\boldsymbol{\theta}}(\mathbf{y}_{n,1:t} \vert D_n, \mathbf{y}_{1:(n-1),1:T})$. We note that
\begin{align}
\label{forward_eval}
p_{\boldsymbol{\theta}}(\mathbf{y}_{n,1:t} \vert D_n, \mathbf{y}_{1:(n-1),1:T})    = \begin{cases}
\frac{p_{\boldsymbol{\theta}}(\mathbf{y}_{n,1:t}, \mathbf{y}_{j:(n-1),1:T} \vert D_n)}{p_{\boldsymbol{\theta}}(\mathbf{y}_{j:(n-1),1:T} \vert D_n)} &\text{if } D_n >1
\\
p_{\boldsymbol{\theta}}(\mathbf{y}_{n,1:t} \vert D_n) \quad &\text{if } D_n = 1
\end{cases}
\end{align}
where $j = \max(1, n-D_n+1)$, can be computed by means of Kalman filters evaluations. Indeed, if $D_n > 1$, 
\begin{align*}
%\label{decomposition1}
    p_{\boldsymbol{\theta}}(\mathbf{y}_{n,1:t}, \mathbf{y}_{j:(n-1),1:T} \vert D_n) =  p_{\boldsymbol{\theta}}(\mathbf{y}_{j:n,1:t} \vert D_n)p_{\boldsymbol{\theta}}(\mathbf{y}_{j:(n-1),(t+1):T} \vert D_n, \mathbf{y}_{j:n,1:t}),
\end{align*}
where $p_{\boldsymbol{\theta}}(\mathbf{y}_{j:n,1:t} \vert D_n)$ is evaluated by a filtering routine up to time $t$ with data of activities with indeces that range between $j$ and $n$, and $p_{\boldsymbol{\theta}}(\mathbf{y}_{j:(n-1),(t+1):T} \vert D_n, \mathbf{y}_{j:n,1:t})$ is evaluated going forward with Kalman filters that treat the element $\mathbf{y}_{n,(t+1):T}$ as missing.
The need to evaluate $p_{\boldsymbol{\theta}}(\mathbf{y}_{j:(n-1),(t+1):T} \vert D_n, \mathbf{y}_{j:n,1:t})$ at any time point requires the ability to perform $T-t$ step ahead Kalman filter evaluations, highlighting the potential computational problem of evaluating the likelihood for long time series (large $T$) and early stages (small $t$). 
One simple solution is to approximate Equation \eqref{forward_eval} with
\begin{align*}
%\label{forward_approx}
p_{\boldsymbol{\theta}}(\mathbf{y}_{n,1:t} \vert D_n, \mathbf{y}_{1:(n-1),1:T})   \propto \begin{cases}
\frac{p_{\boldsymbol{\theta}}(\mathbf{y}_{n,1:t}, \mathbf{y}_{j:(n-1),1:(t+k)} \vert D_n)}{p_{\boldsymbol{\theta}}(\mathbf{y}_{j:(n-1),1:(t+k)} \vert D_n)} &\text{if } D_n >1
\\
p_{\boldsymbol{\theta}}(\mathbf{y}_{n,1:t} \vert D_n) \quad &\text{if } D_n = 1
\end{cases},
\end{align*}
with $k = \min(T-t, k^\star)$ and $k^\star \geq 0$ known, assuming that
\begin{align*}
    \frac{p_{\boldsymbol{\theta}}(\mathbf{y}_{1:(n-1),(t+k+1):T} \vert \mathbf{y}_{1:n,1:t}, \mathbf{y}_{1:(n-1),(t+1):(t+k)},  D_n) }{p_{\boldsymbol{\theta}}(\mathbf{y}_{1:(n-1),(t+k+1):T} \vert  \mathbf{y}_{1:(n-1),1:(t+k)}, D_n)} \propto 1,
\end{align*}
for any $D_n$ and $k$ fixed in advance. This means that whenever a new activity is observed, one needs to simply use a finite number of competing Kalman filters with fixed parameters, where their number is given by the number of different unique particles in the SMC approximation of $p_{\boldsymbol{\theta}}(D_n \vert \mathbf{y}_{1:(n-1),1:T})$.
\addtolength{\textheight}{.5in}%

{In principle, the role of $k^\star$ is to go forward with Kalman filters and to evaluate information that is subsequent to time $t$ but that has already been observed before the $n$-th activity. However, choosing a large $k^\star$ implies the need to proceed with Kalman filter evaluations even many instants after $t$. This could be a problem for contexts in which it is necessary to obtain real-time feedback quickly. A large $k^\star$ allows to go very far ahead with Kalman filter evaluations,  thereby slowing down the computations. Setting $k^\star = 0$ is a practical choice to avoid slowdowns in computations. It is interesting to note that although the information regarding observations after $t$ for activities prior to the $n$-th are not considered by the Kalman filters, they are used in the derivation of $\eta_{n}^B(D_{n})$.}

\section{Simulation studies}
\label{sec:sim}
We investigate here the performance of our proposed changepoint detection algorithm for the between- and within-online settings via a series of simulated data scenarios.  In particular, we illustrate that beyond changepoint identification, and unlike other potentially competitive alternatives \citep[see, e.g. ][]{xie2021sequential}, our methodology can monitor online the probability of a changepoint during the activities.  We fixed  $N = 1000$, $T=60, 120, 200$, $P = 2$, $S= 50$ randomly chosen changepoints and variances  $\sigma_\epsilon^2 = 1$, $\sigma_\alpha^2 = 0.05$, $\sigma_d^2 = 5$, and $\rho = 0.8$.    
With $\boldsymbol{\alpha}_{0}^{(s)} = \mathbf{0}_{2P}$, $\mathbf{\Psi}_0 = \begin{bmatrix} 1/3 & 0.5 \\ 0.5 & 1 \end{bmatrix}$, and $\boldsymbol{\alpha}_{n, 0} = \mathbf{0}_{P}$, 
we generated the shared states for each segment according to 
$
        \boldsymbol{\alpha}_{t+1}^{(s)} = \mathbf{I}_P \otimes \begin{bmatrix} 0.95 & 1 \\ 0 & 0.90\end{bmatrix}  \boldsymbol{\alpha}_{t}^{(s)}$ +$ \boldsymbol{\xi}_t^{(s)}$,  $\boldsymbol{\xi}_t^{(s)} \sim N_{2P}( \mathbf{0}_{2P}, \sigma_\alpha^2 (\mathbf{I}_P \otimes \mathbf{\Psi}_0))$
and the activity-specific states according to 
$       \boldsymbol{\alpha}_{n, t+1} =  {\rho} \cdot  \boldsymbol{\alpha}_{n, t}$ $+ \boldsymbol{\xi}_{n, t}$, $\boldsymbol{\xi}_{n, t}\sim N_{P}( \mathbf{0}_{P}, \sigma_d^2 \mathbf{I}_{P})$.
We then generated the observations 
$    \mathbf{y}_{n,t} = \begin{bmatrix} \mathbf{I}_P \otimes \begin{bmatrix} 1 & 0 \end{bmatrix} & \mathbf{I}_P\end{bmatrix} \begin{bmatrix} \boldsymbol{\alpha}_t^{(s)} \\ \boldsymbol{\alpha}_{n, t} \end{bmatrix}$
$+ \boldsymbol{\epsilon}_{n,t}$,
$\boldsymbol{\epsilon}_{n,t} \sim N_P(\mathbf{0}_P, \sigma_\epsilon^2 \mathbf{I}_P)$.

We set $\lambda = 0.5$ and estimated $\boldsymbol{\theta} = (\sigma_\epsilon^2, \sigma_\alpha^2, \sigma_d^2, \rho)$.   %lthough $\lambda_{\boldsymbol{\theta}}$ can be estimated \added{how? have we written this somewhere? I think you have answered this question before -apologies for re-asking}\added[id = MS]{This is important, we need to discuss better. It's mainly related to \cite{Yildirim2013}, but not used in our paper.} it does not influence other parameters given the high number of activities observations.
In addition, we set $k^\star=0$ since using a small $k^\star$ is a necessary practical choice when $T$ is large.  As $k^\star$ increases, the proposed algorithm for the within-online setting becomes infeasible for large $T$. Alternative specifications of $k^\star$ are investigated %\added{left or investigated?}\added[id = MS]{actually, I have only few trials of it, nothing really relevant to include in the supplementary} 
in the supplementary material together with alternative specifications of $\lambda$.

\begin{figure}
    \centering
    \includegraphics[scale=0.25]{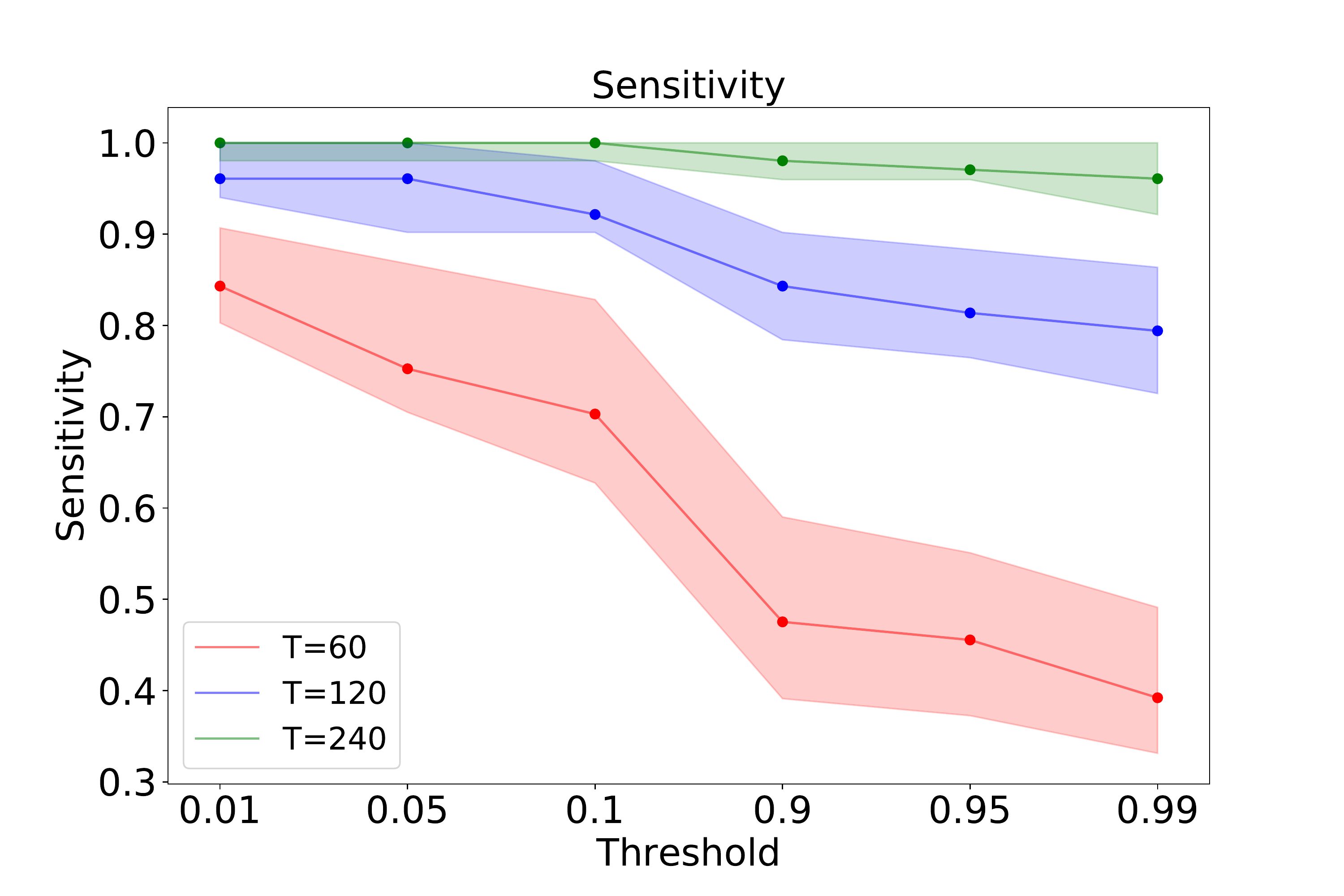}
    \includegraphics[scale=0.25]{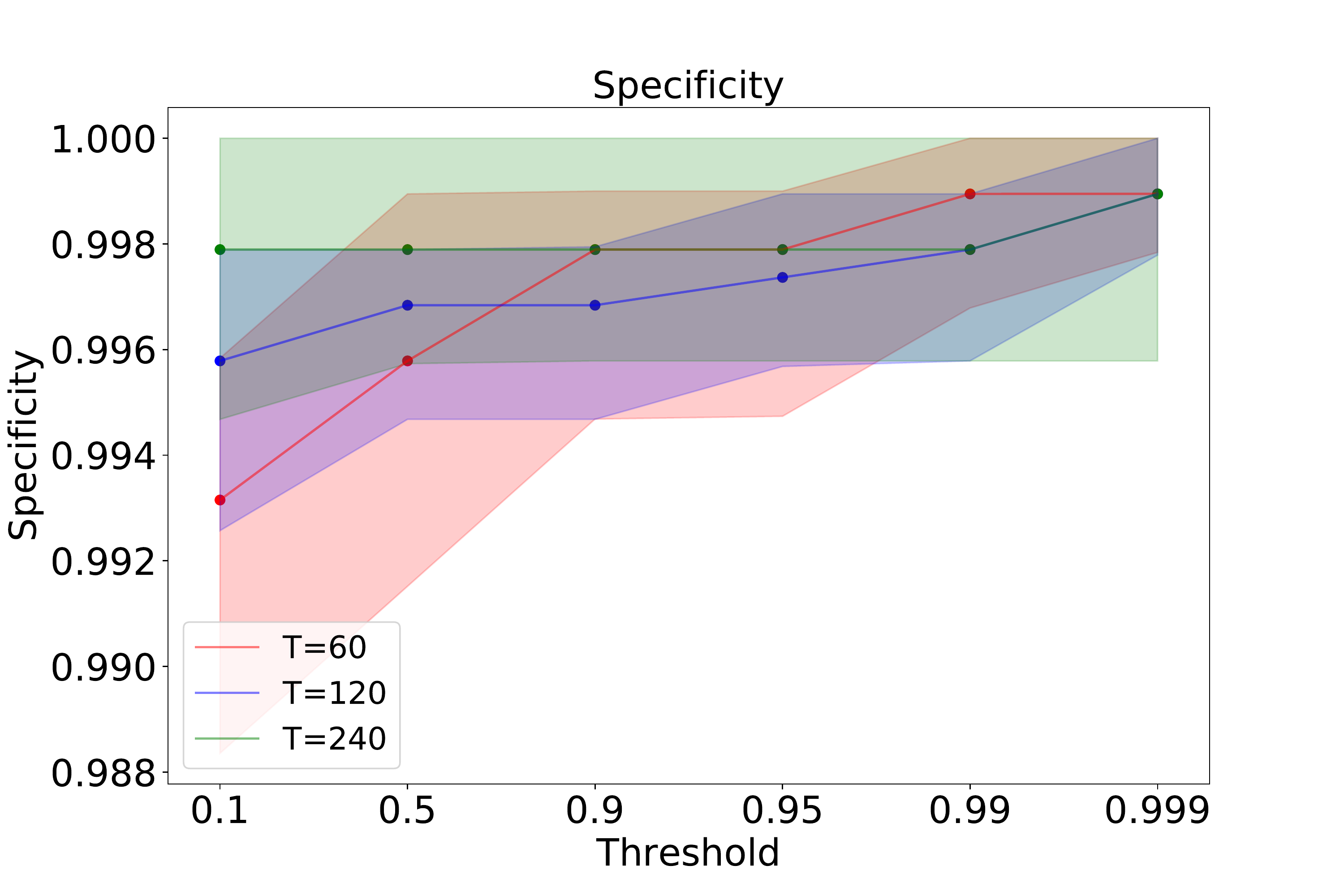}
    \caption{Medians and $90\%$ confidence intervals of sensitivity and specificity with different thresholds obtained for $20$ synthetic examples using our model for $T = 60,120$ and $240$.}
    \label{fig:result_synth_sens}
\end{figure}

We estimate (i) the changepoints in the between-online setting by utilizing Equation \eqref{filtered approx} and testing  
$p_{\hat{\boldsymbol{\theta}}_{1:(n-1)}}(D_n = 1\vert \mathbf{y}_{1:n, 1:T}) > \delta$ for some threshold $\delta$ 
and (ii) the probability of activity $n$ of being a changepoint before it ends in the within-online setting according to $p_{\hat{\boldsymbol{\theta}}_{1:(n-1)}}(D_n = 1\vert \mathbf{y}_{n, 1:t} , \mathbf{y}_{1:(n-1), 1:T})$.
{ Figure \ref{fig:result_synth_sens} depicts the behavior of sensitivity and specificity as the  length of the time series increases, leaving the remaining elements of the models unchanged. We deal with the usual trade-off between  sensitivity and specificity by noting that in our application, maximizing sensitivity ---which is minimizing the number of  activities that are wrongly classified as negative--- is more important, as it may indicate possible activity problems. This is naturally controlled by the threshold $\delta$, see 
Figure \ref{fig:result_synth_sens}.  It is reassuring that our algorithm maintains high levels of specificity as $\delta$ changes, regardless of the length of the time series.  In contrast, the sensitivity seems to decrease significantly as $\delta$ increases, particularly for $T=60$ and $T=120$, although it remains stable for $T=240$.
}

\begin{figure}
    \centering
    \includegraphics[scale=0.23]{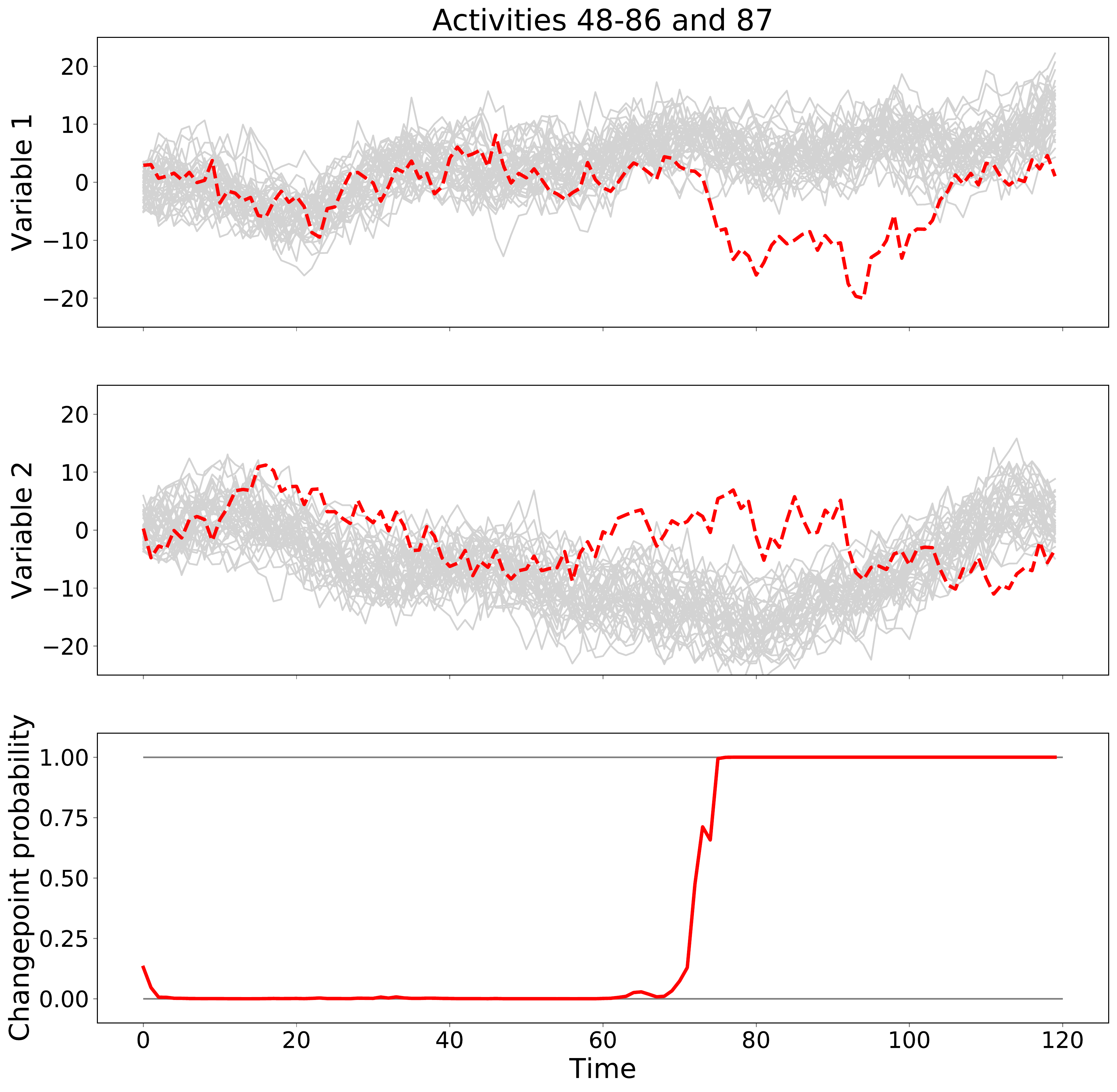}
    \includegraphics[scale=0.23]{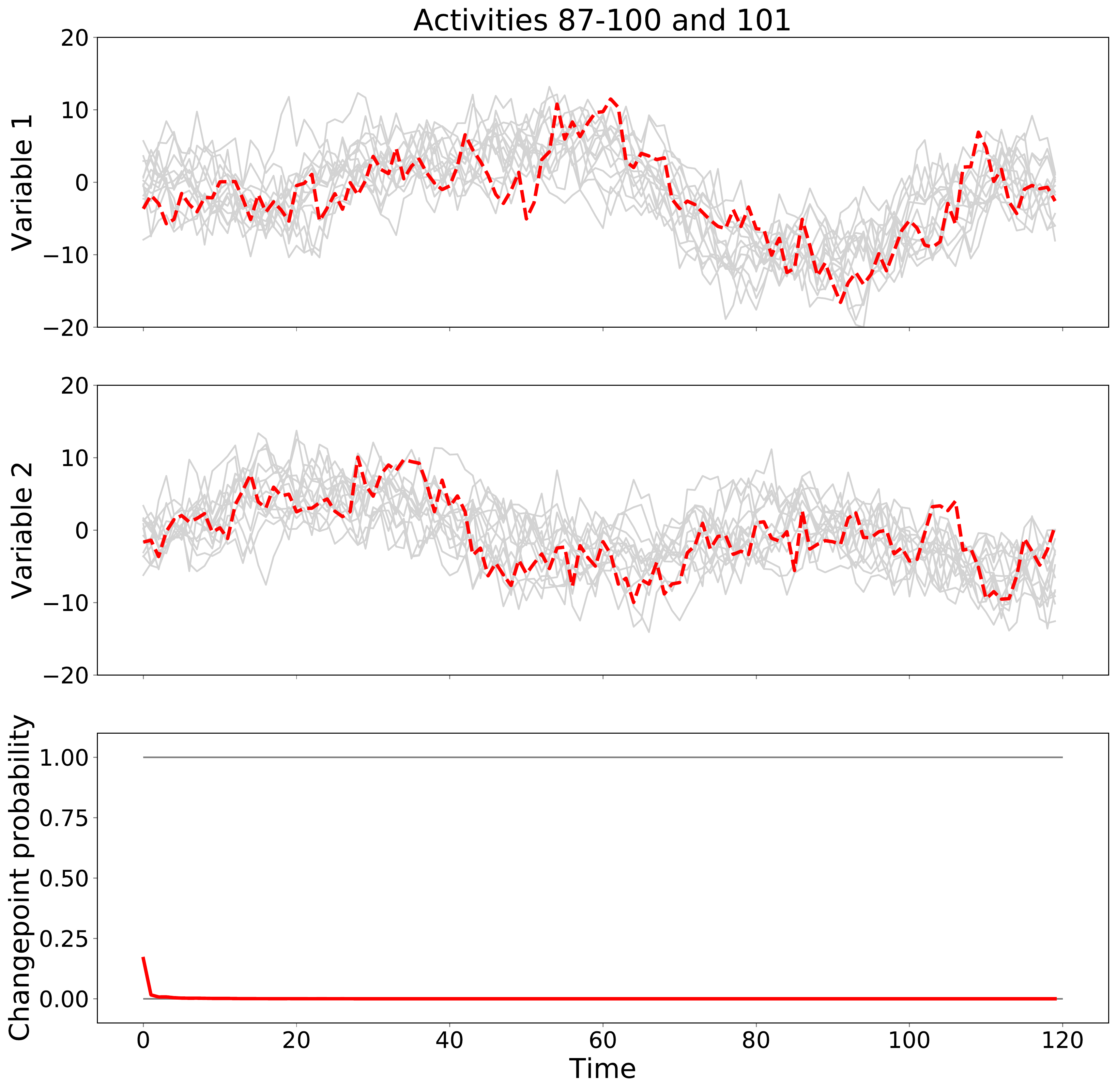}
\caption{Two instantiations of our simulation with different activities and the respective filtered probabilities of changepoints. Red dashed line: activity that is being monitored; gray lines: previous activities since the last changepoint.}
    \label{fig:probability_changepoints}
\end{figure}

\begin{figure}
    \centering
    \includegraphics[scale=0.25]{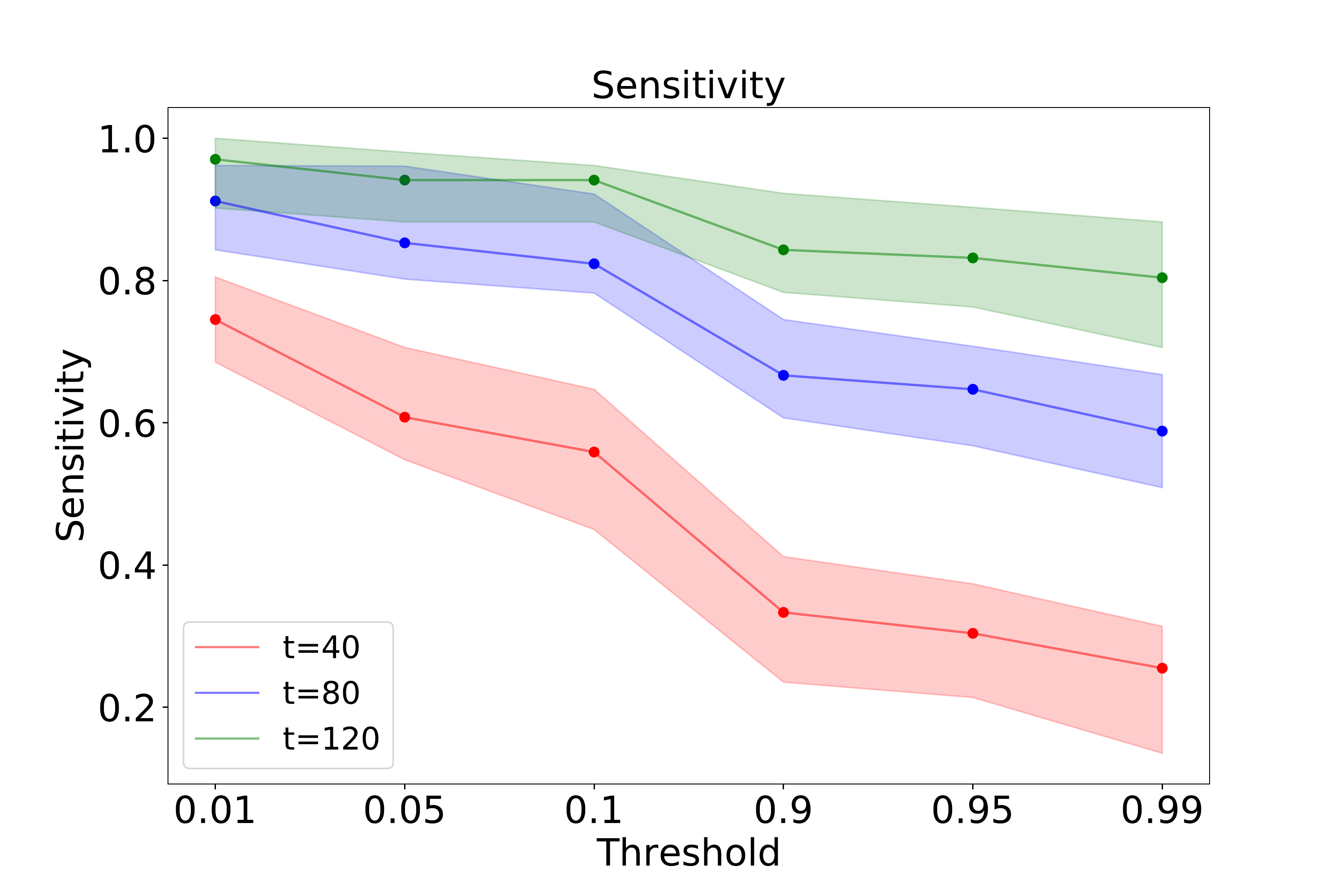}
    \includegraphics[scale=0.25]{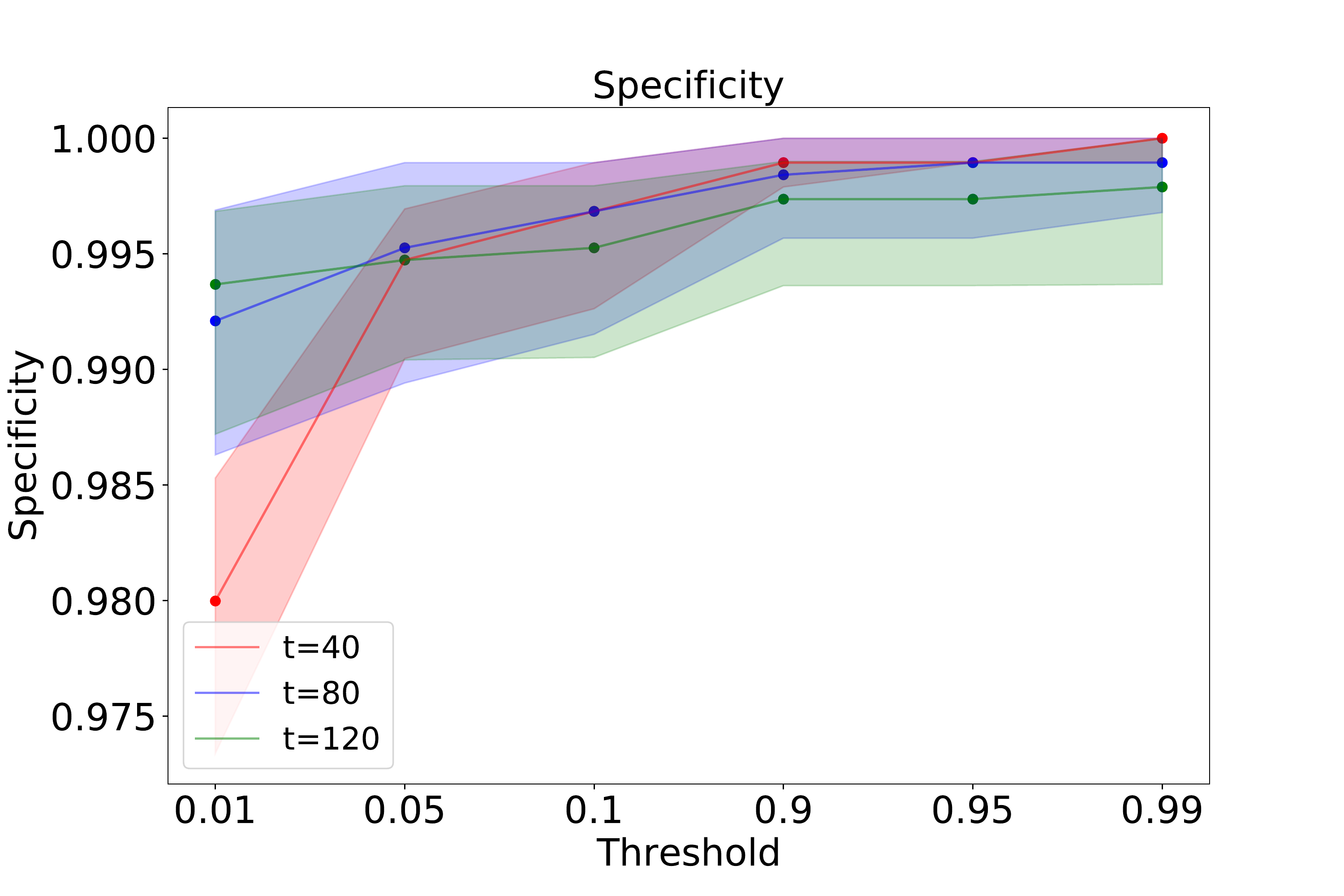}
    \caption{
    Medians and $90 \%$ confidence intervals of sensitivity and specificity evaluated for $20$ synthetic examples using our model in the within-online setting.
}
    \label{fig:sensspec_early}
\end{figure}

The within-online setting allows to monitor online the probability of an activity changepoint, providing information on the athlete's behavior with respect to the past. Figure \ref{fig:probability_changepoints} shows two instantiations of this: the filtered probability \\ $\hat{p}_{\hat{\boldsymbol{\theta}}_{1:(n-1)}}(\mathbf{y}_{n,1:t} \vert D_n, \mathbf{y}_{1:(n-1),1:T})$,  depicted in the bottom row, is estimated online as new observations are collected for the two simulated activities (top two rows). Each panel shows the current (dashed red line) and previous (solid gray) activities since the last changepoint. 
In the within-online setting, the changepoint detection is performed by estimating changepoint probabilities for various values of $\delta$ and $t$ in $20$ replications of the experiment.  Figure \ref{fig:sensspec_early} depicts the results of the simulation study in terms of sensitivity and specificity for different values of $\delta$ and $t$. As expected, the sensitivity drops as $\delta$ increases and as $t$ decreases. Since all time series were simulated with initial values around zero, it is hard to achieve an early (at $t=40$) changepoint detection, although the detection after having observed $2/3$ of the time series (at $t=80$) seems to be satisfactory.

\section{Application}
\label{sec:real}
We consider a set of $85$ warm-up running activities on flat routes consisting of the first $10$ minutes of running of a well-trained athlete. 
The difference between the maximum and minimum altitude reached during each activity was less than $10$ meters, and the activities were measured every second by a Polar v800 smart watch and a Polar H10 heart rate monitor. Warm-up activities are extremely relevant in several sports because they prepare athletes for specific training sessions, influence sports performance, and reduce the risk of injury.  Moreover, they inform on the training status of an athlete just before the training session so early decisions can be made.  In the sports science literature, the choice of the relevant indicators for monitoring the health status and training loads with emphasis on the importance of pre-training analysis is well documented, see, for example,  \cite{buchheit2014monitoring}. In general, heart rate is the most evaluated variable, as it provides insights into oxygen consumption and the physical response to the external stimuli of the exercise \citep{dong2016role, schneider2018heart}.  Heart rate levels during exercise are also influenced by the intensity at which the exercise is performed, represented by the speed of running, which is why we have collected data for both heart rate and speed. Let $y_{\text{hr},n,t}$ be the heart rate in beats per minute and   $y_{\text{sp},n,t}$ be the speed, equal to the difference between the cumulative distances at time $t$ and $t-1$ for activity $n$. We specify a state space model with the measurement equation
\begin{align*}
    \begin{bmatrix}
    y_{\text{hr},n,t}\\
    y_{\text{sp},n,t}
    \end{bmatrix} =    \begin{bmatrix} 1 & 0 & 0\\0 & 0 & 1\end{bmatrix}\begin{bmatrix}
    \alpha_{\text{hr}, 1, t}^{(s)}\\
    \alpha_{\text{hr}, 2, t}^{(s)}\\
    \alpha_{\text{sp}, t}^{(s)}
    \end{bmatrix} + \begin{bmatrix}
    1 & 0  \\ 0 & 1 
    \end{bmatrix} \begin{bmatrix}
    \alpha_{\text{hr}, n, t}\\
    \alpha_{\text{sp}, n, t}
    \end{bmatrix} + \begin{bmatrix}
    \upsilon_{\text{hr}, n,t}   \\
    \upsilon_{\text{sp}, n,t}
    \end{bmatrix}, \quad  \begin{bmatrix}
    \upsilon_{\text{hr}, n,t}   \\
    \upsilon_{\text{sp}, n,t}
    \end{bmatrix} \sim N_2 (0, \mathbf{\Sigma}), 
\end{align*}
segment-specific state equations 
\begin{align*}
\begin{bmatrix}
    \alpha_{\text{hr}, 1, t+1}^{(s)}\\
    \alpha_{\text{hr}, 2, t+1}^{(s)}\\
    \alpha_{\text{sp}, t+1}^{(s)}
    \end{bmatrix}  =    \begin{bmatrix}1 & 1 & 0 \\
    0 & 1 & 0 \\
    0 & 0 & 1
    \end{bmatrix} \begin{bmatrix}
    \alpha_{\text{hr}, 1, t}^{(s)}\\
    \alpha_{\text{hr}, 2, t}^{(s)}\\
    \alpha_{\text{sp}, t}^{(s)}
    \end{bmatrix} +  \begin{bmatrix}
    \xi_{\text{hr}, 1, t}^{(s)}\\
    \xi_{\text{hr}, 2, t}^{(s)}\\
    \xi_{\text{sp}, t}^{(s)}
    \end{bmatrix}, \quad \begin{bmatrix}
    \xi_{\text{hr}, 1, t}^{(s)}\\
    \xi_{\text{hr}, 2, t}^{(s)}\\
    \xi_{\text{sp}, t}^{(s)}
    \end{bmatrix} \sim N_3 (0, \mathbf{\Psi}),
\end{align*}
and activity-specific state equations
\begin{align*}
    \begin{bmatrix}
    \alpha_{\text{hr}, n, t+1}\\
    \alpha_{\text{sp}, n, t+1}
    \end{bmatrix}  = 
    \begin{bmatrix}
    1 & 0  \\ 
    0 & \rho_{\text{sp}}  
    \end{bmatrix}
    \begin{bmatrix}
    \alpha_{\text{hr}, n, t}\\
    \alpha_{\text{sp}, n, t}
    \end{bmatrix} +  \begin{bmatrix}
    \xi_{\text{hr},n,t}
    \\
    \xi_{\text{sp},n,t}
    \end{bmatrix}, \quad  \begin{bmatrix}
    \xi_{\text{hr},n,t}
    \\
    \xi_{\text{sp},n,t}
    \end{bmatrix} \sim   N_2 (0, \mathbf{\Delta}),
\end{align*}
with  $\boldsymbol{\alpha}_1^{(s)} = ( \alpha_{\text{hr}, 1,1}^{(s)}, \alpha_{\text{hr}, 2,1}^{(s)},\alpha_{\text{sp}, 1}^{(s)})^\prime \sim N_3((80,0,0)^\prime, \text{diag}(100,1,100))$, $\boldsymbol{\alpha}_{n, 1} = (\alpha_{\text{hr},n,1}, \alpha_{\text{sp}, n, t})^\prime \sim N_2(\mathbf{0}, 10 \cdot \mathbf{I}_2)$, $\mathbf{\Sigma}$, $\mathbf{\Psi}$, and $\mathbf{\Delta}$ are full covariance matrices, and $\rho_{\text{sp}}$ is an autoregressive coefficient.

The segment-specific latent states that describe the physical condition and skills of the athlete were chosen to be modeled by a linear trend model that captures the segment-specific global trends for the heart rate and by a local level model for the speed.  The activity-specific states are modeled by a random walk process for the heart rate and using an AR(1) process for the speed.
Once the variables are de-trended, the heart rate moves slowly over time, as it does not vary abruptly in healthy conditions, although speed may do so due to, for example, street obstacles.
%added{ I cannot see how the next paragraph is related to AR(1) or random walk}
%We refer with these to aspects that are not directly observed by the smart-watches and that disturb the carrying out of the activity, such as, for example, the need of slowing down due to the presence of a pedestrian or the need of crossing a road. 
We set $\lambda = 0.5$ and, following the guidelines of \cite{Yildirim2013}, we estimated the parameters of the model $\boldsymbol{\theta} = \{  \mathbf{\Sigma}, \mathbf{\Psi},\mathbf{\Delta}, \rho_{\text{sp}}\}$ by repeating the EM algorithm
$30$ times in order to reach convergence. 
%We also include a  random intercept which is fixed  over time in the model to control for the present over-persistence of the Heart Rate variable, which manifests itself through slight level deviations on the level with respect to the segment trend. We externally control the mean and variability of this intercept, setting them respectively at $0$ and $5$, aware of the fact that the level reached by the heart rate is a relevant aspect of the athlete's performance and condition \citep{schneider2018heart}, and therefore that an extreme constant deviation of the Heart Rate has to be considered as a change of behavior.
%Together with controlling the mean and the variance of the intercept , we fix $\rho_{\text{hr}}=0.95$ and $\sigma_{\text{hr}}^{2} = 1$.
\begin{figure}[ht]
    \centering
    \includegraphics[scale = 0.25]{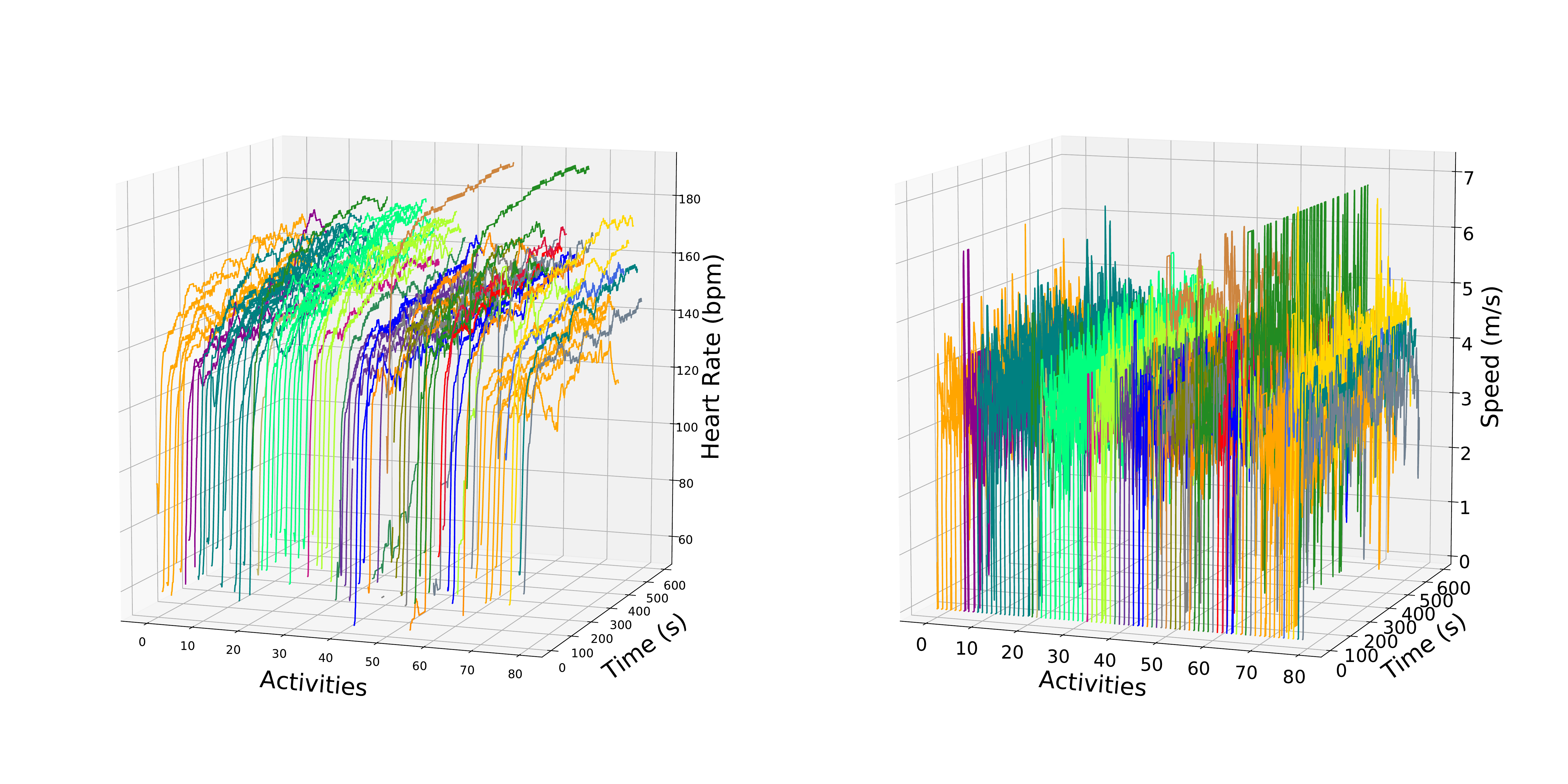}
\caption{Segmentation of warm-up activities in the between-online setting for an athlete. The segmentation of the activities was obtained defining the changepoint as those activities for which the filtered distribution at the end of the activity is  $\hat{p}_{\hat{\boldsymbol{\theta}}_{1:(n-1)}}(D_n \vert \mathbf{y}_{1:n, 1:T})>0.5$. % Time is measured in seconds, heart rate in beats per minute, and speed in meters per second.
%\added{Here and in Figure 7: can you not add this in the plot? eg m/sec, sec, beat/min} \added[id=MS]{I have other plots, different colors}
}
    \label{fig:real_mix}
\end{figure}
Figure \ref{fig:real_mix} provides four instantiations of our results. 
We depict  segments in the between-online setting, obtained according to the rule $\hat{p}_{\hat{\boldsymbol{\theta}}_{1:(n-1)}}(D_n \vert \mathbf{y}_{1:n, 1:T})>0.50$. The estimated number of changepoints is $34$, of which $19$ involve activities with a single activity segment. This interesting finding highlights the large variability between successive activities. Of these $19$ changepoints, $15$ are located in the last $43$ activities and should be attributed not only to changes in the state of the athlete but also to the presence of systematic measurement errors, probably due to a device problem.

\begin{figure}
    \centering
    \includegraphics[scale = 0.23]{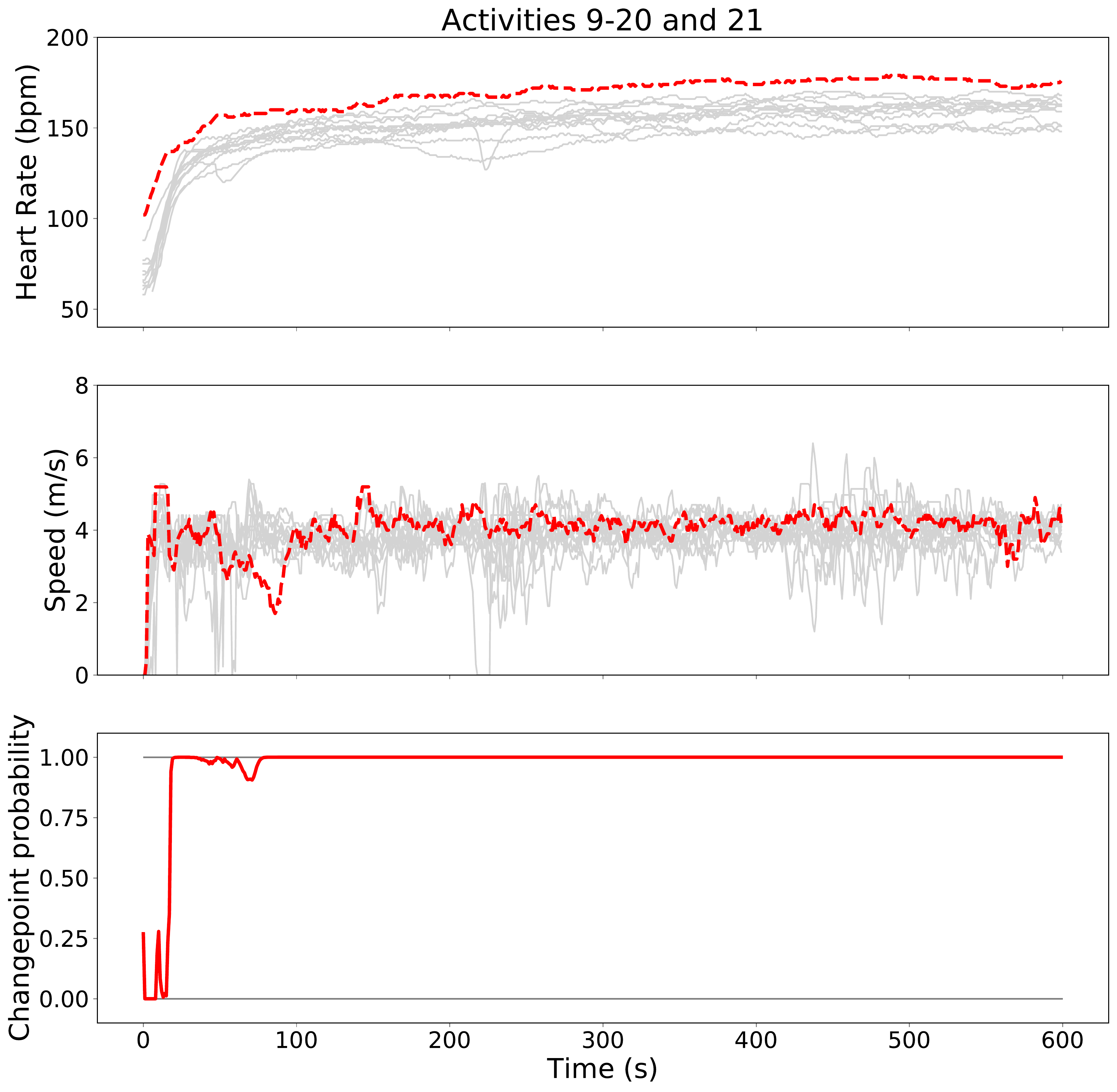}
    \includegraphics[scale = 0.23]{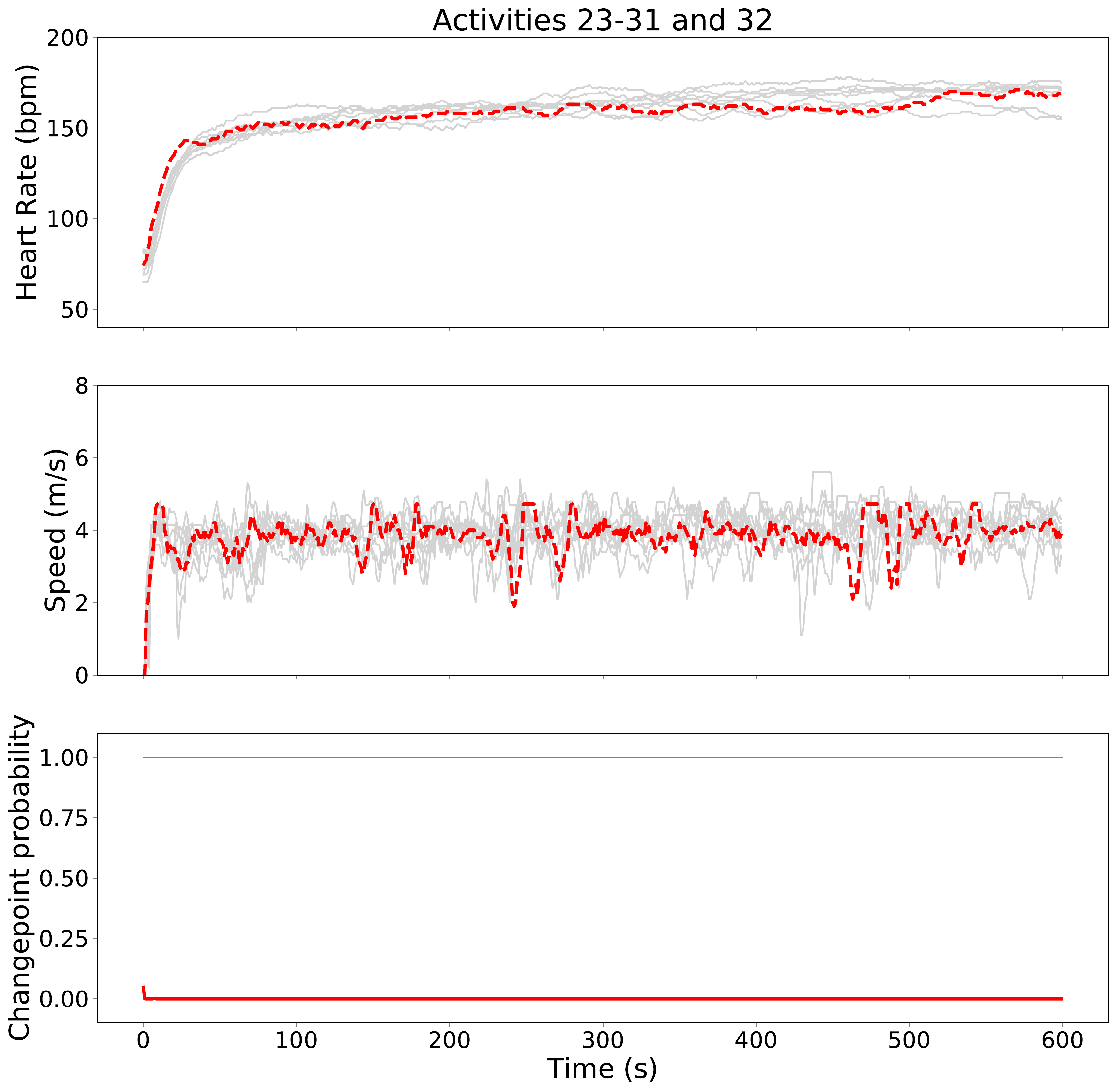}
    \includegraphics[scale = 0.23]{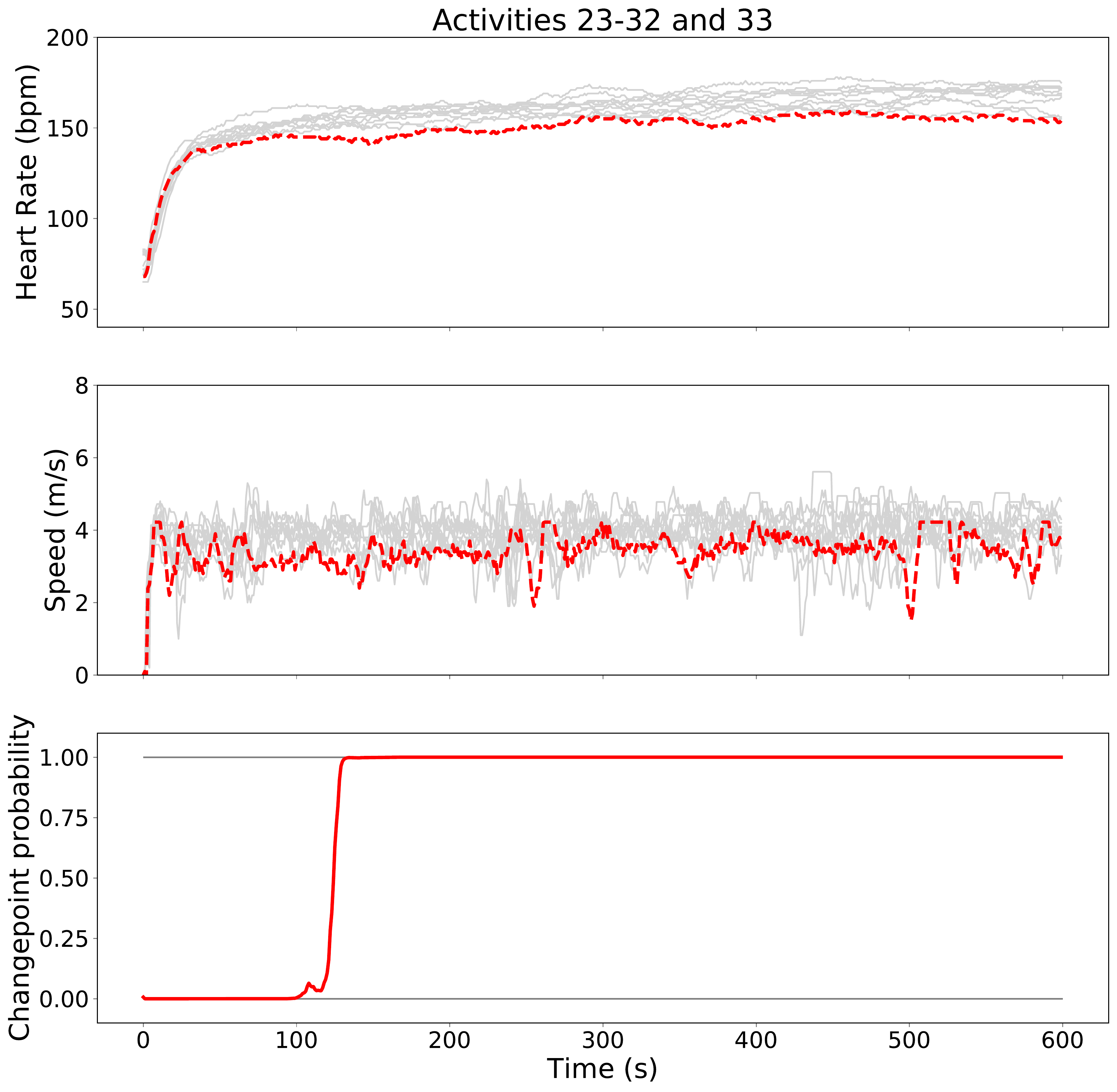}
    \includegraphics[scale = 0.23]{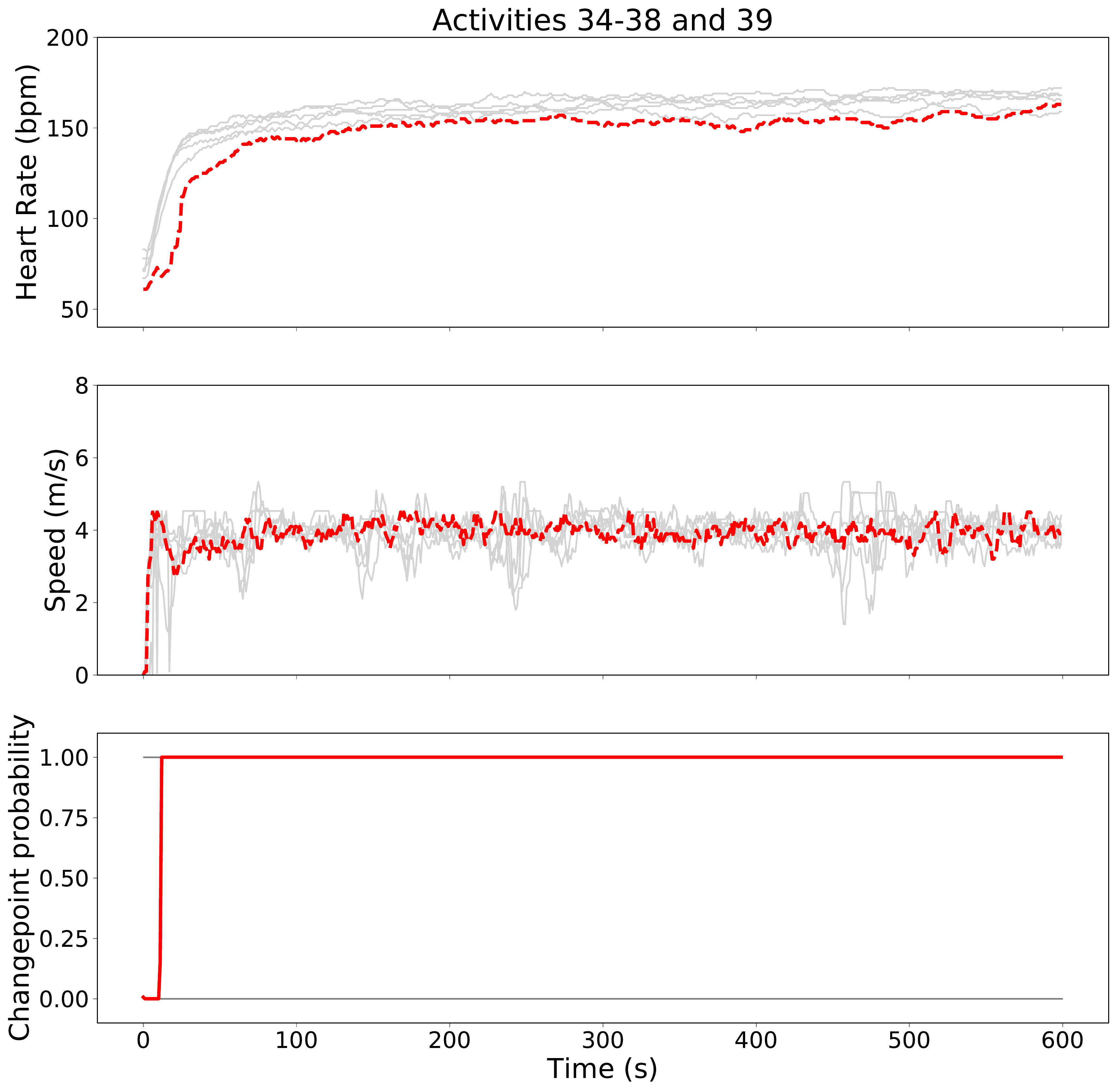}
\caption{Selected activities in which the changepoint probability is being monitored. The gray lines in the background represent activities since the last changepoint, obtained according to the rule  $\hat{p}_{\hat{\boldsymbol{\theta}}_{1:(n-1)}}(D_n \vert \mathbf{y}_{1:n, 1:T})>0.5$.}
    \label{fig:real}
\end{figure}

Figure \ref{fig:real} shows four instantiations of the within-online setting by presenting heart rate, speed, and changepoint probability $\hat{p}_{\hat{\boldsymbol{\theta}}_{1:(n-1)}}(D_n \vert \mathbf{y}_{1:n, 1:T})$ for the monitored activity (dashed red line) and for all activities subsequent to the previous changepoint (solid gray lines) for $\delta = 0.5$. In particular, activity $21$ was identified as a changepoint because a sub-optimal behavior was detected due to a higher heart rate (with similar speed behavior) compared with the previous activities. The changepoint probability is close to $1$ after around $20$ seconds of warm-up. Activity $32$ is similar to activities $23$--$31$ and the changepoint probability is nearly $0$ throughout the activity.  The bottom left panel shows activity $33$, for which the changepoint probability changes strongly after two minutes because both the heart rate and the speed tend to be lower than during other activities, corresponding to athlete putting in less effort. Finally, activity $39$ is characterized by a lower heart rate, although the speed curve seems similar to those in previous activities; this indicates less effort and an improved  state of well-being of the athlete.
\section{Conclusion and future developments}
\label{sec:conclusion}
Motivated by the need to develop an online probabilistic inference framework for runners who collect data using smart devices, we have proposed a new model for changepoint detection in a doubly-online framework.  Our focus lies on the early detection of distributional changes between a set of repeated running activities. The proposed model combines and leverages tools from the classical changepoint model by \cite{Yildirim2013} and the linear and Gaussian state space model \citep{durbinkoopman12,shumway_book}.  The former allows the use of an SMC approach with constant complexity in a between-online framework, while the latter provides the user with updated information on the activity as new data are observed by means of Kalman filter routines.  We adopted a linear and Gaussian state space model, which is a general family of models that allows to include many standard modeling specifications used in time series analysis.  

We considered design matrices that are fixed with respect to both $t$ and $n$ and potentially preclude time-dependent and activity-specific covariates. It is probably reasonable to assume that covariates such as different types of terrain or changes in elevations could affect heart rate or speed. This limitation can be easily overcome by modifying the Kalman recursions appropriately without a substantial change in the remaining methodology. 

We also assumed that activity-specific elements do not interact with segment-specific latent states by imposing a block-diagonal structure on both the transition matrix and the covariance matrix of disturbances in Equation \eqref{s_model_state}. One possible generalization could assume that  the block transition matrix in Equation \eqref{s_model_state} is a block matrix in which the elements outside the diagonal of the first column block are non-zero. This generalization also requires a modification of the Kalman recursions without a substantial change in the methodology, also allowing for activity-specific states determined by some segment-specific states, such as, the autoregressive process with segment-specific coefficients. 

The changepoint prior probability $\lambda$ can be modeled as $\lambda_{\boldsymbol{\theta}} =\lambda_{\boldsymbol{\theta},n} =\lambda_{\boldsymbol{\theta}}(\mathcal{X}_n)$ depending on a set $\mathcal{X}_n$ of time-invariant activity-specific covariates. The set $\mathcal{X}_n$ may represent the meteorological condition during the activity or health-related measures taken prior to the activity, such as heart rate variability in the morning or the number of hours of sleep. This generalization requires the computation of sufficient statistics and a maximization step that is dependent on the specification of the link function.  Both developments are generally related to the standard methods used for the binomial model; see \cite{Yildirim2013} for details. 

A particularly appealing possible future development that requires additional methodological effort is to consider nonlinear and non-Gaussian state space models. This might be of interest in contexts in which the use of smart devices allows for the collection of varied data \citep{bourdon2017monitoring}, violating the common Gaussian assumptions.

\bigskip
\begin{center}
{\large\bf SUPPLEMENTARY MATERIAL}
\end{center}

Please write {mattia.stival@unipd.it} for supplementary material, including proofs, details, data, and derivation.

\bibliographystyle{agsm}
\bibliography{Biblio.bib}
\end{document}